\documentclass[11pt]{amsart}
\usepackage{amsaddr}

\usepackage[margin=1in]{geometry}                
\usepackage{graphicx}
\usepackage{amssymb}
\usepackage{epstopdf}
\usepackage{url,hyperref,lineno,microtype,subcaption}
\usepackage[onehalfspacing]{setspace}
\usepackage{graphicx,color,amsmath,subeqnarray}
\usepackage{soul, xcolor}

\newcommand{\bc}{\begin{center}}
\newcommand{\ec}{\end{center}}
\newcommand{\beq}{\begin{equation}}
\newcommand{\eeq}{\end{equation}}
\newcommand{\benum}{\begin{enumerate}}
\newcommand{\eenum}{\end{enumerate}}
\newcommand{\bea}{\begin{eqnarray*}}
\newcommand{\eea}{\end{eqnarray*}}
\newcommand{\beqa}{\begin{eqnarray}}
\newcommand{\eeqa}{\end{eqnarray}}
\newcommand{\bal}{\begin{align}}
\newcommand{\eal}{\end{align}}
\newcommand{\ba}{\begin{array}}
\newcommand{\ea}{\end{array}}
\newcommand{\nonum}{\nonumber}
\newcommand{\bi}{\begin{itemize}}
\newcommand{\ei}{\end{itemize}}
\newcommand{\vs}{\varepsilon}
\newcommand{\sigE}{\sigma_E}
\newcommand{\sigF}{\sigma_F}
\newcommand{\sigS}{\sigma_S}
\newcommand{\del}{\delta}
\newcommand{\ts}{t^{\textrm{spike}}_Q}
\newcommand{\spee}{S_E^Ep_E^E}
\newcommand{\spei}{S_I^Ep_I^E}
\newcommand{\speF}{S_{E}^Fp_{E}^F}
\newcommand{\speS}{S_{E}^Sp_{E}^S}
\newcommand{\spiF}{S_{I}^Fp_{I}^F}
\newcommand{\spiS}{S_{I}^Sp_{I}^S}

\newcommand{\odor}{\textrm{odor}}
\newcommand{\red}{\textcolor{red}}

\renewcommand{\th}{\textrm{th}}

\begin{document}

\title{Network Mechanism for Insect Olfaction}
\author{Pamela B.~Pyzza\,$^{1}$, Katherine A.~Newhall\,$^{2}$, Gregor Kova\v ci\v c\,$^{3}$, Douglas Zhou\,$^{4}$, and David Cai\,$^{4,5}$}
\date{\today}                                           

\address{$^{1}$Department of Mathematics and Statistics, Kenyon College, Gambier, OH, USA \\
$^{2}$Department of Mathematics, University of North Carolina at Chapel Hill, Chapel Hill, NC, USA\\
$^{3}$Department of Mathematical Sciences, Rensselaer Polytechnic Institute, Troy, NY, USA\\
$^{4}$Department of Mathematics, MOE-LSC, and Institute of Natural Sciences, Shanghai Jiao Tong University. Shanghai, China\\
$^{5}$Courant Institute of Mathematical Sciences, New York University, New York, NY, USA}

\date{Received: date / Accepted: date}

\maketitle

\begin{abstract}
Early olfactory pathway responses to the presentation of an odor exhibit remarkably similar dynamical behavior across phyla from insects to mammals, and frequently involve transitions among quiescence, collective network oscillations, and asynchronous firing.  We hypothesize that the time scales of fast excitation and fast and slow inhibition present in these networks may be the essential element underlying this similar behavior, and design an idealized, conductance-based integrate-and-fire (I\&F) model to verify this hypothesis via numerical simulations.  To better understand the mathematical structure underlying the common dynamical behavior across species, we derive a firing-rate (FR) model and use it to extract a slow passage through a saddle-node-on-an-invariant-circle (SNIC) bifurcation structure.   We expect this bifurcation structure to provide new insights into the understanding of the dynamical behavior of neuronal assemblies and that a similar structure can be found in other sensory systems. 
\end{abstract}

\section{Introduction}

Despite major differences in the details of both the olfactory neuronal network architecture and its odor response even among insects~\cite{Carcaud2016,Barbara:2005rw,MacLeod:1996th,Ng:2002ij,Tanaka:2009uq,Heinbockel:1998ys}, 
early olfactory pathways appear to share a number of universal anatomical and functional characteristics across animal phyla ranging from insects to mammals~\cite{Hildebrand:1997kx,Eisthen:2002fk,Kay:2006ea,Laissue:2008pd}.   Partly due to these shared characteristics, a significant effort has been expended on experimentally studying insects as animal models of olfaction~\cite{Sato:2009zr,Joerges:1997hc,Christensen:2000fu,Tanaka:2009uq}.   In particular, the antennal lobe (AL) in the insect plays the role analogous to that of the mammalian olfactory bulb~\cite{Kay:2006ea} as the first brain area in which the neuronal computations involving the olfactory stimuli take place that reformat these stimuli, before transmitting them to further downstream brain areas~\cite{Hildebrand:1997kx,STRAUSFELD1999634,Eisthen:2002fk,Chen2005,Ache:2005uq,Kay:2006ea,WACHOWIAK2006411}.    Correspondingly, a widely shared functional characteristic among insects and mammals is the frequent presence of fast, synchronized, collective oscillations in the dynamics taking place in these two analogous areas~\cite{Kay:2006ea,Sivan:2006fy,KAY2015}.

The similarities in the AL network connectivity architecture among several frequently-studied insect species~\cite{MacLeod:1996th,Ng:2002ij,Tanaka:2009uq,Heinbockel:1998ys,Carcaud2016,Barbara:2005rw}  and the somewhat similar manner in which synchronized oscillations appear and disappear after the odor presentation in most if not all of these species~\cite{Turner2008,Tanaka:2009uq,Laurent:1996bh,Stopfer:1997ve,Heinbockel:1998ys}, hint at a possible \emph{robust}, \emph{universal}, \emph{network} mechanism that may underlie the temporal evolution of these oscillations.
In this paper, we hypothesize one such mechanism, which is  the interaction among different time scales of excitation and inhibition in the AL neuronal network~\cite{Wilson:2005kb,Barbara:2005rw,Grunewald:2003fu}.  We also identify a mathematical structure that can be used to describe this mechanism. 

As an example, we investigate this mathematical structure for the locust, which possesses a simple AL architecture and its corresponding dynamics, and has been studied especially thoroughly both experimentally and theoretically~\cite{MacLeod:1996th,Laurent:1994fu,Laurent:1996fv,Laurent:1993ye,Laurent:1994dz,Bazhenov:2001ai,Bazhenov:2001pd,PRC09,PRC2011,Sivan:2006fy}.   The hypothesized network architecture and dynamical scenario that emerge from these studies are as follows:  The locust AL network is composed of excitatory projection neurons (PNs) and inhibitory local neurons (LNs), which are believed to communicate through fast excitatory (2--5 ms) and fast (30--40 ms) and slow (200--300 ms) inhibitory currents~\cite{Laurent:1993ye,Bazhenov:2001pd,Bazhenov:2001ai}.   (The presence of the slow inhibitory current in the locust AL has been inferred indirectly~\cite{MacLeod:1996th,Stopfer:1997ve,MacLeod:1998rr,Bazhenov:2001pd,Bazhenov:2001ai,Sachse:2002aa,Barbara:2005rw,PRC09}, see, e.g., the explanation in ref.~\cite{PRC09}.)
The PNs generate action potentials while the LNs generate long, 20-30~ms calcium spikes~\cite{MacLeod:1996th,Bazhenov:2001ai,Bazhenov:2001pd}.   Upon receiving the stimulus from the olfactory receptor neurons (ORNs) that reside on the antennae, the AL network dynamics proceed in three stages:  First the network begins generating collective oscillations with frequency~$\sim 20$~Hz~\cite{Laurent:1994fu,Laurent:1993ye,Heinbockel:1998ys}, detectable from the local field potential, which is related to the average excitatory neuron voltage over the network~\cite{Laurent:1994dz,Laurent:1996lq}.  Second, a brief quiescent period~\cite{Laurent:2001tg} follows.   Third,  the neurons' firing rates modulate slowly in a manner determined by the stimulus for about 1 second until they reach a steady state~\cite{Wehr:1996xr}, and, after the odor subsides, the excitatory neuronal firing rates again exhibit such slow modulation while they settle back into baseline equilibrium during the next few seconds~\cite{Mazor:2005eu}. 

Two complementary theoretical hypotheses for odor-encoding mechanisms have emerged from the studies described in the previous paragraph.  The first is \emph{temporal binding}~\cite{SG95,Malsburg:1999fe,Christensen:2000fu,Lei:2002bv}, expressed by a group of neurons that consistently participate in all the cycles of the initial AL network oscillations within about 500~ms from the stimulus onset~\cite{PRC2011}.  These oscillations appear to be generated by the interaction between the excitation and fast inhibition, and damped by the growing presence of slow inhibition.   The second is \emph{slow patterning}, the collective time dependence of the firing-rate trajectories traced out by the excitatory neurons in the AL over 2--4 seconds,  generated by the interplay between the excitation and slow inhibition in the AL~\cite{Bazhenov:2001ai,Bazhenov:2001pd,Stopfer:2003kl}. Based on these time scales, the temporal binding mechanism takes place during the first, collective oscillations, stage of the dynamics, which suggests that it helps the insect detect a brief plume of a transient odor.   The slow patterning mechanism, in turn, takes place during the third stage of the dynamics, which suggests that it may sharpen the identification of a persistent odor~\cite{PRC09,PRC2011}.

To test our hypothesis that the dynamical evolution of the locust AL dynamics emerging upon odor presentation results primarily from the interaction among the excitatory and two inhibitory response time-scales, in contrast to the previous detailed Hodgkin-Huxley-type  point-neuron models~\cite{Bazhenov:2001ai,Bazhenov:2001pd,Sivan:2004bs,Sivan:2006fy,PRC09,PRC2011}, we follow the parsimonious approach outlined in ref.~\cite{RTKC09b} and begin our modeling by employing the more idealized, conductance-based integrate-and-fire (I\&F) model~\cite{Burkitt06a,Burkitt06b}.    To be able to use the I\&F model in the description of the locust AL dynamics, we must choose the rise and decay time scales of the stereotyped postsynaptic conductance responses in such a way that they reflect both the presynaptic input and synaptic receptor time courses.     We then proceed with our investigation in three steps: In the first step, we find that an appropriately inhomogeneously-driven, sparsely-coupled I\&F network reproduces both the above-described three-stage dynamical scenario, including its evolution time scales, as well as the dynamical behavior conjectured to underlie the hypothesized temporal binding and slow patterning odor-encoding mechanisms. In particular, this I\&F model is able to capture odor discriminability via each of these mechanisms.  Thus, our model confirms the presence of a robust \emph{network}, as opposed to an intrinsic neuronal, mechanism underlying the  AL dynamics. 

In the second step, in search for a mathematical structure underlying the locust AL dynamics, we embark on a progression of further idealizations.  We first apply an idealized \emph{white odor}, which is assumed to drive all excitatory and all inhibitory neurons in the network uniformly.  
We then replace sparse network coupling with all-to-all network coupling and thus also idealize the network architecture, so that we arrive at a network with no specific choice of either the input signal or the network connectivity. 
With each of these subsequent idealizations, we find that the three-stage dynamical scenario not only persists but becomes more pronounced in that the initial oscillations become better synchronized and the quiescent period less noisy, with this period still followed by an interval of slowly-modulated neuronal firing rates that reach a steady state before the odor subsides.   This result points to the three-stage scenario as forming a structurally-robust feature underlying the locust AL network dynamics, arising from the interaction between the fast excitation and the fast and slow inhibition.

 In the third step,  we idealize the model yet further by coarse-graining the dynamics of the aggregate excitatory and fast and slow inhibitory network conductance responses in the AL stimulated with a white odor-like input using a four-dimensional, slow-fast \emph{firing-rate} (FR) model~\cite{Tre93,SM02,SMSW02,cai2006} in order to reveal a simple mathematical structure underlying the robust, three-stage  scenario of AL network dynamics.  As expected, this FR model again reproduces this scenario, and thus confirms the leading underlying physiological mechanism to be the interaction between the fast excitation and the fast and slow inhibition.  This model also lets us identify a corresponding mathematical structure describing the three-stage scenario as a slow passage through a \emph{saddle-node-on-an-invariant-circle} (SNIC) bifurcation.   

The remainder of the paper is organized as follows.  In Section~\ref{METHODS}, we present the I\&F model, diagnostic tools that we use to highlight and quantify the model's ability to reproduce experimentally observed dynamics, and a heuristic derivation of the four-dimensional FR model (with details in Appendix~\ref{apdx:firingratemodel}).  In Section~\ref{INSECT_OLFACTION}, we confirm that our  I\&F model reproduces the three-stage dynamical scenario observed in experiments, different slow patterns of neuronal firing-rate trajectories as a result of different model odor stimuli, and the presence of temporally-bound neurons that discriminate among the odors.    In Section~\ref{IDEALIZATIONS}, we submit the point-neuron model to further idealizations, in the last one of which the network is all-to-all coupled, mean-driven, and receives a white odor stimulus. This network allows for coarse graining and retains the  three-stage dynamical scenario.  In Section~\ref{BIFURCATIONS}, we describe how the FR model reveals the mathematical structure underlying the network mechanism as a slow passage through a SNIC bifurcation.  Conclusions and further discussion can be found in Section~\ref{DISCUSSIONS}. The details of the FR model derivation can be found in Appendix~\ref{apdx:firingratemodel}. Details concerning the robustness of the method we use for determining the sets of temporally-bound neurons participating in network oscillations, which we use to discriminate among odors, are described in Appendix~\ref{apdx:BIvalidity}.   A linearized version of the FR model is presented in Appendix~\ref{apdx:linearization}. Its derivation is presented in Appendix~\ref{apdx:linearizedmodel} and the presence of a unique limit cycle in this linearized model is verified in detail in Appendix~\ref{apdx:limitcycle}.


\section{Methods}\label{METHODS}

\subsection{Integrate-and-Fire Model\label{IFmodel}}
We build a model of $N$ conductance-based, integrate-and-fire (I\&F) point neurons with two time scales for inhibitory postsynaptic responses.  The evolution of the membrane potential of the $i^{\th}$ neuron in the network, $v_i(t)$ for  $i=1,\dots, N$, is governed by the equation
\begin{equation}\begin{aligned} 
\tau \frac{dv_i(t)}{dt} =& -\left[v_i(t) - \vs_R\right] - g_i^E(t)\left[v_i(t) - \vs_E\right] \\
& - g_i^F(t)\left[v_i(t) - \vs_F\right] - g_i^S(t)\left[v_i(t) - \vs_S\right],    \label{voltage}
\end{aligned}\end{equation}
whenever $v_i(t)$ is below the firing threshold, $V_T$. Here, $\tau$ is the membrane time constant, $g_i^E(t)$, $g_i^F(t)$, and $g_i^S(t)$ are the neuron's time-dependent excitatory, fast inhibitory, and slow inhibitory conductances, respectively, and $\vs_R$, $\vs_E$, $\vs_F$, and $\vs_S$, are the reversal potentials corresponding to the leakage, excitatory, fast inhibitory, and slow inhibitory conductances, respectively, with $\vs_S < \vs_F < \vs_R < V_T < \vs_E$. The event when $v_i(t)$ reaches the firing threshold, $v_i(t) = V_T$, represents the neuron firing an action potential.  We do not model this action potential explicitly, but rather reset $v_i(t)$ to $\vs_R$,  hold it there for a refractory period of 5~ms,  and update the postsynaptic conductances according to the rule described below.  

In the locust, the excitatory neurotransmitter is acetylcholine~\cite{Koch99}.  A detailed model of the corresponding receptor response is used in refs.~\cite{Bazhenov:2001pd,Bazhenov:2001ai,PRC09,PRC2011}. In our idealization, and with an eye on further coarse graining our model, we instead model the postsynaptic excitatory-response shape using the stereotypical form of $t e^{-t/\sigma_E}$.  What we do need to capture accurately, however, is the time scale of the acetylcholine receptor, which is still achieved using the stereotypical form of our conductance response. Thus, we use the excitatory conductance decay rate $\sigma_E = $ 1--2~ms, corresponding to the excitatory time scale of 2--4~ms~\cite{PRC09}. (Note that in the response form $te^{-t/\sigma}$, the true response time is longer than the decay rate $\sigma$.) This stereotyped form is widely used, for example, in the modeling of the mammalian primary visual cortex~\cite{SNS95,MSSW00,Burkitt06a}, and is similar to that generated by an AMPA receptor \cite{Johnston:1983uq}. 

The inhibitory neurons generate calcium-dependent ``spikes"~\cite{Laurent:1996fv}, which, in contrast to action potentials, are prolonged increased levels of activity, lasting approximately 20~ms. Nonetheless, in our I\&F model network, we give up on accurately modeling the inhibitory neurons' membrane potentials, and focus on capturing the correct time scale of the postsynaptic inhibitory conductance response.  We thus again model the fast inhibitory conductances by the stereotypical form $t e^{-t/\sigma_F}$, adjusting its decay rate to match the corresponding correct time scale and therefore we take $\sigma_F=$ 4--8~ms.   This amounts to a lengthening of the time scale of the fast conductance response as compared to the true GABA$_A$ receptor response time of 3--5~ms~\cite{PRC09}, to account for the duration of the calcium spike.  (Again, note that the response time scale is longer than the decay constant, $\sigF$.)

As mentioned in the Introduction, the presence of a slow, inhibitory current with a response time of several hundred milliseconds has been inferred indirectly, as described in ref.~\cite{PRC09}.    (The corresponding GABA$_B$ receptors have been found in the honey bee and moth~\cite{Grunewald:2003fu,Barbara:2005rw,Heinbockel:1998ys}, which have olfactory systems similar to that of the locust.)  Thus, as in refs.~\cite{Bazhenov:2001ai,Bazhenov:2001pd,PRC09,PRC2011}, we include a slow current in our model.   To conform with the rest of our modeling approach, we drive this current by  
a stereotyped slowly-changing, inhibitory conductance response of the form 
$(e^{-t/\rho_S}-e^{-t/\sigma_S})/(\rho_S-\sigS)$~\cite{TKPM98,Burkitt06a},  with rise time scale $\rho_S=$ 420--500~ms and decay time scale $\sigS = $ 600--800 ms, that is initiated concurrently with the fast inhibitory conductance response. (A more realistic slow conductance response is described in refs.~\cite{Bazhenov:2001ai,Bazhenov:2001pd,PRC09,PRC2011}).  

The stereotyped excitatory, fast inhibitory, and slow inhibitory conductance responses for the $i^{\th}$ neuron are governed by second-order kinetics, described by the equations

\begin{subeqnarray} \label{conductance}
\sigma_P \frac{dg_i^P(t)}{dt} &=& -g_i^P(t) +h^P_i(t), \slabel{geqn}\\
\sigma_E \frac{dh_i^E(t)}{dt} &=& -h_i^E(t) + f_{i}^{\odor} \sum_{l} \del(t - \tau^i_l) + f_{i} \sum_{k} \del(t - \gamma^i_k) \nonumber \\
&& + \frac{S^E_{i}}{N_E}\sum_{j \neq i}p^E_{ji}\sum_{\mu}\del(t-t^j_{\mu}), \slabel{heqn} \\
\sigma_F \frac{dh_i^F(t)}{dt} &=& -h_i^F(t) + \frac{S^F_{i}}{N_I}\sum_{j \neq i}p^F_{ji}\sum_{\mu}\del(t-t^j_{\mu}), \slabel{heqn2} \\
\rho_S \frac{dh_i^S(t)}{dt} &=& -h_i^S(t) + \frac{S^S_{i}}{N_I}\sum_{j \neq i}p^S_{ji}\sum_{\mu}\del(t-t^j_{\mu}), \slabel{heqn3} 
\end{subeqnarray}
with $P = E$, $F$, and $S$ and $\delta(\cdot)$ the Dirac-delta function. 
The synaptic strengths are encoded in the coefficients $S^E_i$, $S^F_i$ and $S^S_i$; they each take one of two values depending on the $i^{\th}$ neuron's type (E or I), and are scaled by the size of the corresponding population ($N_E$ or $N_I$) in anticipation of future coarse graining over large networks.   We keep the ratio $N_E/N_I=3$ constant, as has been observed in the locust AL \cite{Leitch:1996ly}. 

The parameters $p^E_{ji}$, $p^F_{ji}$, and $p^S_{ji}$ are the elements of three different network adjacency matrices: $p^P_{ji} = 1$ (recall $P = E$, $F$, and $S$) if  the  corresponding type of synaptic connection is present from neuron $j$ to neuron $i$, and $p^P_{ji}=0$ otherwise. The network is constructed by randomly choosing synaptic connections between pairs of neurons with the following probabilities: $p^E_{E}$ and $p^E_I$ for excitatory connections to excitatory and inhibitory neurons, respectively, $p^F_{E}$ and $p^S_{E}$ ($p^F_{I}$ and $p^S_{I}$) for fast and slow inhibitory connections to excitatory (inhibitory) neurons.  Experimental evidence points to the fact that the fast and slow inhibitory receptors are colocalized~\cite{Barbara:2005rw,Enell2007,LECORRONC2002419,Cayre1999}, so we take $p^F_{ji}=p^S_{ji}$, and also $p^F_{Q}=p^S_{Q}$, $Q=E$ or $I$,  in the rest of the paper.  

The times $t_\mu^j$ appearing in the Dirac-delta functions of Eq.~\eqref{conductance} represent the $\mu^{\textrm{th}}$ spike time of the $j^{\textrm{th}}$ neuron.  At such times, the corresponding $h_i^P$ of the postsynaptic neurons jump by the prescribed amounts, $S_i^P/N_Q$, $Q=E$, or $I$.

Equation~\eqref{conductance} also includes external-drive spikes from both an odor-specific source with synapse strength $f_i^{\odor}$ and a background source with synapse strength $f_i$.  
We model the presentation of an odor by 
driving a subset of neurons (typically 1/3 of them) with a set of excitatory external spikes at times $\tau_l^i$, generated by a set of independent Poisson processes with common rate $\nu^{\odor}$ and synapse strength $f_E^{\odor}$ or $f_I^{\odor}$,  depending on the $i^{\th}$ neuron's type ($E$ or $I$).   The external spikes drive the excitatory conductances on both the excitatory and inhibitory neuronal populations.  
It is experimentally known that both excitatory PNs and inhibitory LNs receive input from the olfactory receptor neurons \cite{Barbara:2005rw,Kay:2006ea}, so neither $f_E^{\odor}$ or $f_I^{\odor}$ vanishes.


The background noise drives the excitatory conductances of all neurons with spikes of strength $f_{i}$,  taking one of two values, $f_E$ or $f_I$, depending on the $i^{\th}$ neuron's type ($E$ or $I$).  We model the arrival times of the background spikes, $\gamma_{k}^i$, to each neuron with independent Poisson processes, all with an identical rate, $\nu$.  

In our simulations, as listed in Table~\ref{Tbl:LargeParameterList}, we use the frequency $\nu=4000$ Hz for the excitatory background drive, and choose the corresponding synaptic strength $f_E$ such that the overall excitatory background-drive magnitude is $f_E\nu=8$.  This leads to the membrane potentials of the neurons in the AL network being primed just below the firing threshold, with a possible rare firing, agreeing with refs.~\cite{PRC09,Perez-Orive:2002nx}.    Likewise, we use the frequency $\nu_{\odor} = 6000$ Hz for the odor drive, which represents the convergent input to a PN or LN from about 200 ORNs firing at about 30 Hz~\cite{PRC09}.   We  choose the excitatory and inhibitory strengths $f_E^{\odor}$ and $f_I^{\odor}$ of the spikes arriving from the ORNs such that the overall excitatory and inhibitory drive magnitudes become $f_E^{\odor}\nu^{\odor}=6.9$ and $f_I^{\odor}\nu^{\odor}=6.6$, respectively.    Together, all of these parameter choices induce the initial network oscillations to occur at the frequency of about 20 Hz.   Moreover, choosing $f_E^{\odor}\nu^{\odor} > f_I^{\odor}\nu^{\odor}$ ensures that the LNs are excited and begin firing when PNs begin firing, but do not fire sooner than most PNs would and thus suppress the PN firings during the initial oscillatory period.

The model is normalized such that $\vs_R = 0$, $\vs_E = 14/3$, $\vs_F = -2/3$, $\vs_S = -9/5$, and $V_T = 1$. We integrate the model equations numerically using an algorithm developed in ref.~\cite{ST01}, which employs the second-order Runge-Kutta method for integrating the potential equation between neuronal spikes, linear interpolation to find the spike times and reset the membrane potential, and an exact solution for the conductance equations. 


\subsection{Diagnostic Tools}\label{ss:diagnostic}

Here we describe the tools we use to confirm that our I\&F model captures the dynamical phenomena observed experimentally in the AL.    We use the same tools as in refs.~\cite{Mazor:2005eu,PRC09,PRC2011} for ease of comparison. 

\subsubsection{Power Spectral Density (PSD)} 
We use the PSD to determine the dominant frequencies at which the model network oscillates. We also compute the time dependence of the instantaneous power contained in several frequency windows~\cite{gardiner2004handbook,vetterling2002numerical}.  We measure network activity as the average voltage across all of the excitatory neurons in the network, which is used to roughly model  the local field potential (LFP) measured experimentally~\cite{Laurent:1996bh,PRC09}.  We thus refer to this average excitatory neuron voltage as the LFP of the simulation.  We compute the PSD as the magnitude of the discrete Fourier transform calculated using \textsc{MatLab}'s built-in \verb fft   routine (which is based on the FFTW algorithm~\cite{FFTW}) over a given time window of the LFP data.

We compute the amount of power in a given frequency band as a function of time by applying a moving time window to the LFP data.   Specifically, we move a 300~ms time window in increments of 50~ms over the LFP and extract the amount of power in three dominant frequency bands: 6 to 14~Hz, 16 to 24~Hz, and 26 to 34~Hz, by integrating the PSD over each of these frequency bands. We then average this total power of the PSD within each frequency band over 20 simulations of the dynamics. The error of this sample mean is computed as the sample standard deviation divided by the square-root of the number of simulations, $\sqrt{20}$ in this case.  In each simulation, we vary the realization of the external drive to the network but hold the network architecture fixed. 

\subsubsection{Binding Index (BI)} \label{sec:bidef}


The value of the BI, introduced in ref.~\cite{PRC2011}, is used to determine odor-specific triplets of excitatory neurons (PNs) that with high probability fire together during each cycle in the initial oscillatory phase of the network dynamics. The binding index is defined as 
\[BI_{i,j,k} = \min(P_{j,k|i}, P_{i,j|k}, P_{i,k|j}),\]
 where $P_{j,k|i}$ is the conditional probability that excitatory neurons $j$ and $k$ spike, given that excitatory neuron $i$ spikes. Numerically, $P_{j,k|i}$ is calculated by creating a 20~ms window around each spike time of neuron $i$ and determining if neurons $j$ and $k$ both spike within that window. The number of times this occurs divided by the total number of times neuron $i$ spikes is our estimate of the conditional probability $P_{j,k|i}$. For a given $i,j,k$ triplet (order non-specific), if $BI_{i,j,k} \geq b$, where $b$ is a specified threshold, then we consider the triplet temporally bound.

\subsubsection{Principal Component Analysis (PCA)} 

We use  PCA~\cite{PCA} to visualize the dynamics of the entire network in a reduced three-dimensional space, highlighting the odor-specific trajectory that the I\&F dynamics traverse after odor presentation. (Cf. refs.~\cite{Mazor:2005eu,PRC09}.) For a fixed network structure, we created three odors by stimulating three different subsets of 1/3 of the neurons.  For each odor, we ran 50 realizations of the dynamics and sorted the firing times of each of the PNs into 50~ms bins.   We applied \textsc{MatLab}'s \verb PCA  function to this trial-averaged matrix of spike-count data to extract the first three dominant components.   


\subsection{Firing-Rate Model}\label{sec:firingratemodel}

To better understand the mathematical structure underlying the dynamics of the I\&F model, we derive a firing-rate (FR) model for the idealized case of an all-to-all coupled AL neuronal network driven by the ``white-odor" stimulus, i.e., a stimulus that drives all the neurons at an equal rate and with equal strength.
In this FR model, we replace the detailed neuronal voltages and spike trains generated using the I\&F model by the time-dependent neuronal firing rates, $m_E(t)$  and $m_I(t)$, for the typical excitatory and inhibitory neurons, respectively.  Thus, we reduce the dynamics to those of two populations of statistically equivalent neurons, such that 
all the neurons in a population are equally driven when an odor is presented, i.e., there is a uniform external-drive term for each population in the FR model.  

The FR model becomes a closed set of equations via the realization that the population firing rates, $m_E(t)$  and $m_I(t)$,  drive the typical neuronal conductances, $g_Q^E(t)$, $g_Q^F(t)$ and $g_Q^S(t)$, where $Q=E$ or $I$  stands for the excitatory or inhibitory population, respectively, while these conductances feed back into the equations for the firing rates.  The FR model can be reduced further by observing that the conductances $g_E^P$ and $g_I^P$, $P=E$, $F$, or $S$, are not independent, but are instead connected via the relations
\begin{subequations}\label{eqn:substitutions}
\begin{eqnarray}
g^E_Q &= &f_Q\nu+S^E_Q g^E, \label{sub_gE} \\
g^F_Q &= &S^F_Qg^F, \label{sub_gF}\\
g^S_Q &= &S^S_Qg^S, \label{sub_gS}\\
h_Q &= &S^S_Qh,  \label{sub_h}
\end{eqnarray}
\end{subequations}
in which we refer to the common quantities $g^E$, $g^F$, and $g^S$ as the  \emph{effective conductance variables}.
The resulting FR model is a system describing these effective conductances, given by the equations 
\begin{subequations}\label{4Dsystem}
\begin{eqnarray}
  \sigE \frac{dg^E}{dt} &=& -g^E+m_E(t), \slabel{4eqn_gE} \\
  \sigF \frac{dg^F}{dt}&=& -g^F+ m_I(t), \slabel{4eqn_gF} \\ 
  \sigS\frac{dg^S}{dt}&=& -g^S + h, \slabel{4eqn_gS} \\   
\rho_S \frac{dh}{dt}&=& -h + m_I(t), \slabel{4eqn_h} 
\end{eqnarray}
\end{subequations}
where the firing rates $m_E$ and $m_I$ are expressed in terms of the effective conductances as 
\begin{subequations}\label{firing_rates}
\beq\label{meqns_simple}
m_Q(t) =\frac{1+f_Q\nu+S^E_Q g^E+S^F_Q g^F+S^S_Q g^S}{\tau \ln\left[M_Q\right]} ,
\eeq
\beq\label{denom_fr}
M_Q  = \frac{ f_Q\nu(\vs_E - \vs_R)+\sum_{P\in\{E,F,S\}} S^P_Q g^P(\vs_P - \vs_R)}{\left\{ \Delta\vs_R   +f_Q\nu\Delta\vs_E+\sum_{P\in\{E,F,S\}}S^P_Q g^P\Delta\vs_P   \right\}^+ },
\eeq
with $Q=E$ or $I$.
\end{subequations}
Here, $\{ x \}^+=x$ if $x>0$, and zero otherwise, and 
\begin{equation}
\Delta\vs_Z = \vs_Z - V_T\text{ for }Z\in\{ R,E,F,S\}. \label{deltaz}
\end{equation} 
The conditions for which the firing rates in Eqs.~\eqref{firing_rates} do not vanish are given by the inequalitites
\beq\label{boundarycondition2again}
 \Delta\vs_R   +f_Q\nu\Delta\vs_E+\sum_{P\in\{E,F,S\}}S^P_Q g^P\Delta\vs_P > 0, \quad Q=E~\mbox{or}~I. 
 \eeq
When the denominator of the expression $M_Q$ in Eq.~\eqref{denom_fr} vanishes, the corresponding firing rate, $m_Q(t)$ in Eq.~\eqref{meqns_simple}, is set to zero.


\section{Results}

We first demonstrate that our I\&F model is capable of accurately capturing the firing patterns observed in insect olfaction~\cite{Laurent:1996bh,Mazor:2005eu,Friedrich:2001zr}.  In Section~\ref{INSECT_OLFACTION} we show our model network's ability to discriminate among different odors in a manner consistent with the two neural-code mechanisms conjectured in detailed models~\cite{Bazhenov:2001pd,Bazhenov:2001ai,PRC09,PRC2011}, as discussed in the Introduction. Then, in Section~\ref{IDEALIZATIONS}, we focus on the robust underlying feature of the dynamics --- a three-stage progression of activity characterized by 20~Hz oscillations, followed by quiescence, and then followed by slow patterning, being preserved through a sequence of idealizations ending with an all-to-all coupled I\&F network driven by a ``white-odor" stimulus.  Finally, in Section~\ref{BIFURCATIONS}, we show that the FR model resulting from coarse graining this last, most idealized I\&F network model still preserves the three-stage dynamical progression. Moreover, we find this progression in the FR model to be the result of an underlying SNIC bifurcation.

\subsection{Odor Response and Discriminability\label{INSECT_OLFACTION}}

While the I\&F model does not represent the true time course of the inhibitory calcium spikes, it is nonetheless designed to accurately capture the dynamics of the network conductances and firing patterns of the inhibitory neurons. Therefore, as we demonstrate computationally in this section, it indeed captures the network firing-pattern behavior observed experimentally in insect olfaction~\cite{Laurent:1996bh,Mazor:2005eu}.
In particular, our model excitatory and fast-inhibitory currents produce the initial 20~Hz oscillations in the local field potential.  During this initial oscillatory stage, we also find odor-specific, temporally bound PNs, which may be used in odor identification~\cite{PRC2011}.   These initial dynamics are followed by a short quiescent period, and that is followed by slow, odor-specific patterning of the neuronal firing rates~\cite{Bazhenov:2001ai,Bazhenov:2001pd,PRC09}.   Both of these last two stages of dynamics are at least in part induced by the model's slow inhibitory currents.
 
\begin{figure}[ht]
 \begin{centering}
\includegraphics[width = 0.9\textwidth]{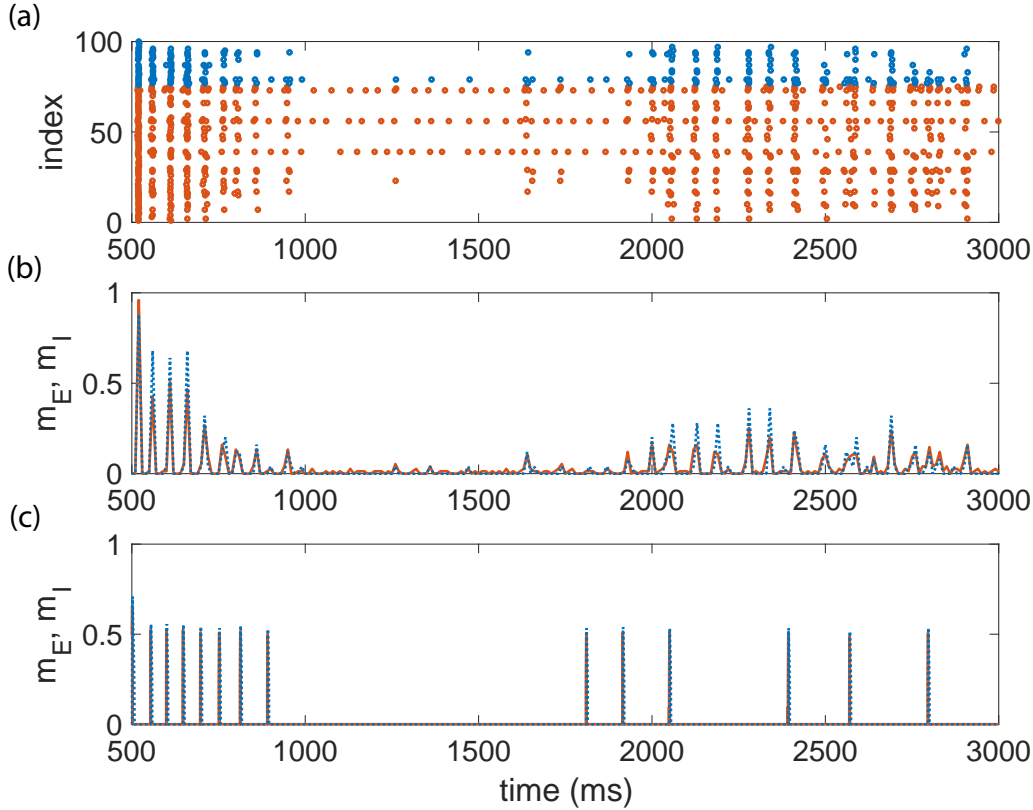}
\caption{(a) Raster plot of excitatory (red, index $1\to N_E=75$) and inhibitory (blue, index $76\to N_E+N_I=100$) populations.  (b) The average per-neuron firing rate over 10~ms bins of the firing pattern in panel (a) for the excitatory (red solid line) and inhibitory (blue dotted line) populations  (c) The evolution of the FR model in Eqs.~\eqref{4Dsystem} and \eqref{firing_rates}, showing $m_E$ (red solid line) for the excitatory and $m_I$ (blue dotted line) for the inhibitory population firing rate. 
Parameters for all panels can be found in Table~\ref{Tbl:LargeParameterList}.}
\label{fig:raster_fire_cond}
\end{centering}
\end{figure}

Our model reproduces these three stages as shown in the the raster plot of Fig.~\ref{fig:raster_fire_cond}(a) when presented with an odor at time $500$~ms. When an odor is presented to a subset of neurons in the model, their voltages are driven towards threshold, with the bulk of the excitatory neurons reaching it slightly ahead of the bulk of the inhibitory neurons when the stimulus drives the excitatory neurons harder than the inhibitory neurons ($f_E^\textrm{odor} > f_I^\textrm{odor}$), and also because the time scale of the excitatory response is faster than that of the inhibitory response ($\sigma_E<\sigma_F$).  The initial excitatory firing recruits other neurons into a synchronous firing event, in which the inhibitory neurons fire, shutting down for a time all neuronal activity due to the elevated values of the fast inhibitory conductances before the firing resumes again.  This is the basic principle underlying the pyramidal-interneuronal network gamma (PING) like oscillation \cite{WHITTINGTON2000,Borgers:2005tv}.  

While these oscillations take place, the slow inhibitory conductances are also building up.  However, their slowly-rising nature results in the effects of the slow inhibition building up over several synchronous PING-like firing events.  At the conclusion of that build-up, almost all of the firing is shut down by the slow inhibitory currents.    Subsequently, as the excitatory membrane potentials recover, the network settles into a new regime of firing, which involves all three conductances. The three main stages of the above I\&F model dynamics are similarly captured by the FR model in Eqs.~\eqref{4Dsystem} and~\eqref{firing_rates}, as shown in  Fig.~\ref{fig:raster_fire_cond}(c).

The synchronization of network oscillations due to the interaction of the excitation and fast inhibition, and the subsequent desynchronization due to the interaction between the fast excitation and slow inhibition, were already studied using the quadratic I\&F model in ref.~\cite{Martinez:2008pi}.   Here, we use our I\&F model network to also reproduce the types of network dynamics conjectured to underlie the two odor-encoding mechanisms described in the Introduction~\cite{Bazhenov:2001ai,Bazhenov:2001pd,Stopfer:2003kl,PRC09,PRC2011}, and exhibit odor discriminability.

\begin{table}[ht]
\renewcommand{\arraystretch}{1.5}
\caption[Parameter Values for Most Figures]{Parameters used for most figures. Some parameters do not pertain to all simulations (e.g. $\nu$ for Fig.~\ref{fig:raster_fire_cond}c) and are therefore left without a value in the table. The number denoted by $\dagger$, indicates the same network connections were used for both fast and slow synapses, i.e., the fast and slow inhibitory receptors are taken to be colocalized.  Note coupling constants should not be directly compared between the two columns because the FR model (Eqs.~\eqref{4Dsystem}) assumes all-to-all coupling while the I\&F model (Eq.~\eqref{IFmodel}) has connectivity generated by the indicated probabilities.}
 \label{Tbl:LargeParameterList}
\begin{center}
\begin{tabular}{c|c|c|}
\multicolumn{1}{c}{} & \multicolumn{2}{|c|}{{\bf Figure Number}}\\ \hline
\multicolumn{1}{c|}{{\bf Parameter}} &  \ref{fig:raster_fire_cond}a, \ref{fig:raster_fire_cond}b, \ref{fig:PSD}, \ref{fig:odor_raster}, \ref{fig:odor_PCA}, \ref{fig:networkbif} & \ref{fig:raster_fire_cond}c, \ref{fig:gSvaries}, \ref{fig:bifurcation}, \ref{fig:EFastI_gIlinearization_withplanes}   \\ \hline
$N_E$ 					& 75 				& -- 				 								\\
$N_I$ 					& 25				& --												\\
$\nu$ 					&4000\text{ Hz }	& --						\\
$f_E\nu$ 					& 8				&  0.3											\\
$f_I\nu$ 					& 0				& 0.024										\\
$\nu^{\odor}$ 				&6000\text{ Hz }	& --						\\
$f_E^{\odor}\nu^{\odor}$ 		& 6.9				& --											\\
$f_I^{\odor}\nu^{\odor}$ 		& 6.6				& --											\\
$\sigma_E$ 				& 1\text{ ms }		& 4.5\text{ ms }						\\
$\sigma_F$ 				& 4\text{ ms }		& 18\text{ ms }						\\
$\rho_S$ 					& 420\text{ ms }	& 420\text{ ms }			\\
$\sigma_S$ 				& 800\text{ ms } 	& 800\text{ ms } 		 	 \\
${S^E_E}$ 					& 6			&0.363									\\
${S^E_I}$ 					& 23.62 		 	& 0.45 		 	 		 	 		 	\\
${S^F_E}$ 					& 43.75 		 	& 5.0		 	 		 	 		 	\\
${S^F_I}$ 						& 8.75		 	& 0.003		 			 			 	\\
${S^S_E}$ 					& 78.75			& 5.0									 \\
${S^S_I}$ 					& 15.75			& 0.003									\\
$p^E_E$ 					& 0.13		& -- \\
$p^E_I$ 					& 0.07		& -- \\
$p^F_E$ 					& 0.15			& -- \\
$p^F_I$ 					& 0.72			&-- \\
$p^S_E$ 					& 0.15$\dagger$			& -- \\
$p^S_I$ 					& 0.72$\dagger$			& -- \\
  \hline
\end{tabular}
\end{center}
\end{table}

\begin{figure}[t]
 \centering
\includegraphics[width = \textwidth]{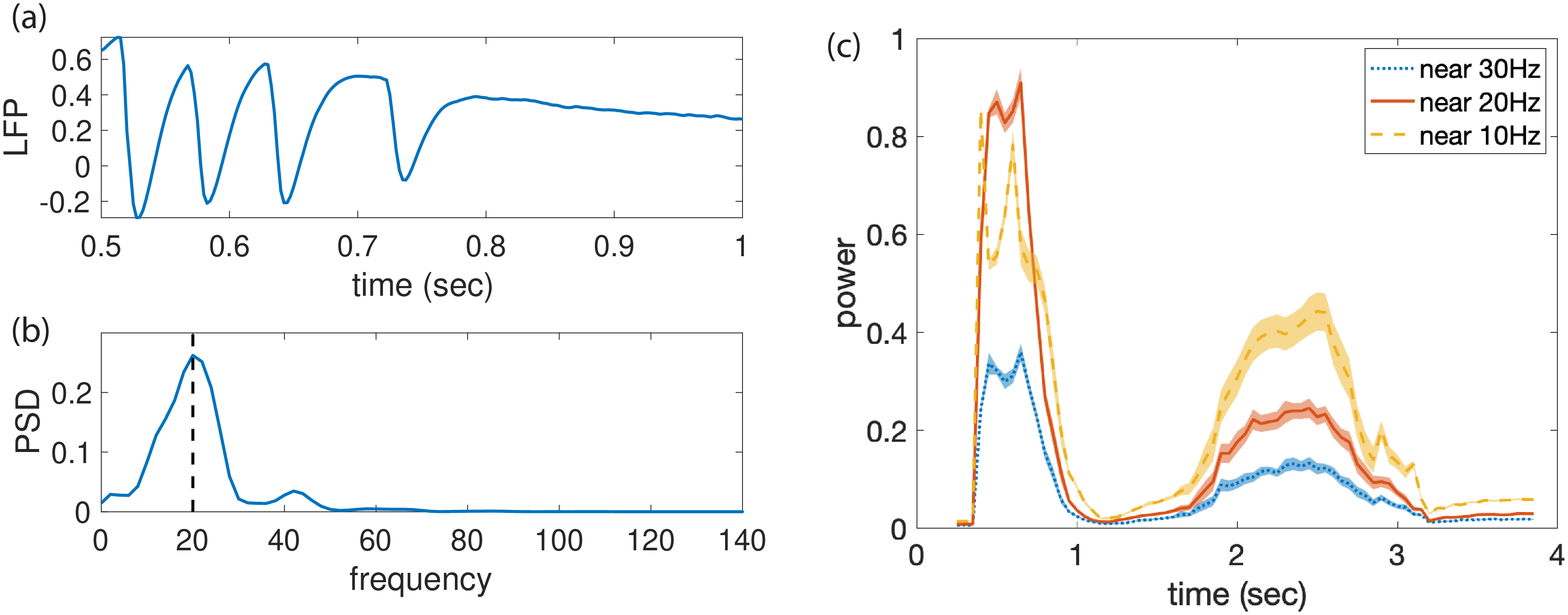}
\caption{(a) The network LFP over a window of 500~ms starting from odor onset.  (b) The PSD for the time period indicated in (a), showing a peak near 20~Hz.  (c) The window-averaged value of the PSD within the ranges of 6 to 14~Hz, 16 to 24Hz, and 26 to 34~Hz, averaged over 20 simulations, as a time window of length 300~ms is moved across the LFP time course in 50~ms steps.  The shaded region indicates the error of the presented sample mean.  The parameters used to generate the LFP are the same as those in Fig.~\ref{fig:raster_fire_cond}(a), which can be found in Table~\ref{Tbl:LargeParameterList}.}
\label{fig:PSD}
\end{figure}

Thus, to further explore the 20~Hz oscillations, at least initially exhibited by both our I\&F and FR models, we compute the I\&F network LFP as the average voltage across all excitatory neurons, shown in Fig.~\ref{fig:PSD}(a) for the first 500~ms of time after odor onset. The PSD of this 500~ms LFP window is shown in Fig.~\ref{fig:PSD}(b), exhibiting the peak frequency near 20~Hz.  In Fig.~\ref{fig:PSD}(c) we show how the power in this 20~Hz peak changes as the dynamics evolve in time.  As described in Section~\ref{ss:diagnostic}, we move a 300~ms time window across the LFP in 50~ms steps, computing the PSD of the 300~ms long window at each step. We integrate the power within each prescribed frequency band, 6--14~Hz, 16--24Hz, and 26--34~Hz, for each window location. Finally, we average this power over 20 realizations and plot the average PSD power versus the center of the time window in which it was calculated, shown in Fig.~\ref{fig:PSD}(c). The results indicate that the 20~Hz network oscillations are strongest during the initial 0.5 seconds of odor presentation, followed by a clear decrease in all activity between 1 and 2 seconds, before activity, characterized by oscillations in the LFP with a broad peak around 10~Hz in the associated PSD, resumes.

Within the initial oscillatory stage, we find temporally-bound neurons specific to the set of odor-driven neurons.  Such bound neurons have been proposed in ref.~\cite{PRC2011} as a possible mechanism of encoding the odor. The temporally-bound neurons fire together with high probability during each oscillation cycle.  To identify these neurons, following ref.~\cite{PRC2011}, we compute the binding index (BI), a measure for triplets of excitatory neurons that equals 1 if these neurons always fire together and never independently, and zero if they always fire independently.  We classify temporally-bound neurons as those belonging to triplets with $BI_{i,j,k} \geq b$, where $b = 0.65$. This value is chosen in agreement with the results in ref.~\cite{PRC2011}.

\begin{figure}[ht]
 \centering
 \includegraphics[width = 0.45\textwidth]{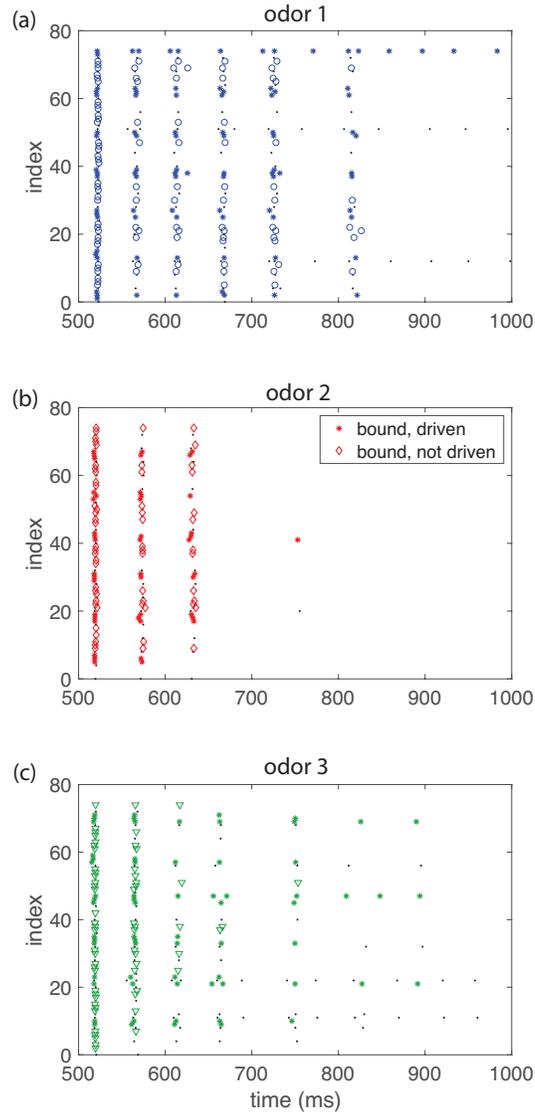}
\caption{
Raster plots of  the excitatory neurons in a network to which three different odors are presented, generated by stimulating a different 1/3 of the neurons in the network. Temporally-bound neurons are those with binding index, $BI_{i,j,k} \geq 0.65$.  In all panels, black dots represent unbound neurons, stars represent temporally-bound neurons directly driven by the odor, and open symbols represent temporally bound neurons not directly driven by the odor. Parameters for all panels are found in Table~\ref{Tbl:LargeParameterList}.} \label{fig:odor_raster}
\end{figure}

In Fig.~\ref{fig:odor_raster}, we display the raster plots of the network response to three different model odor stimuli, each represented by the stimulation of a different 1/3 of the neurons in the network.   The spike times of the bound neurons are indicated by the colored symbols within these raster plots, while the black dots represent the spike times of the unbound neurons.  Of particular interest are the neurons that participate in the collective oscillations and become temporally bound  due to the network dynamics  and are not driven directly by the stimuli arriving from the ORNs, as these are believed to be involved in the early recognition of the odor~\cite{PRC2011}.  They are indicated by open symbols. The neurons that participate in the collective oscillations due to the direct stimulus arriving from the excited ORNs are indicated by stars.

From Fig.~\ref{fig:odor_raster}, we can glean the data displayed in Table~\ref{Tbl:BIdata065}.   This data show that each model odor in our network induces the oscillations of about 30 temporally-bound model neurons that are not directly driven by the ORNs, of which about a half are shared with another odor and the other half are not.  Thus sufficient numbers of such neurons exist so that they can be used to discriminate between pairs of odors.  Moreover, for each odor, at least one indirectly-driven, temporally-bound neuron belongs to the set of the temporally-bound neurons oscillating due to that particular odor and no other, and is thus associated uniquely to that odor.   This shows that the set of the temporally-bound neurons for each odor in our model also provides a unique odor identifier.  (The robustness of the binding-index-threshold choice in light of these results, in particular, the robustness of our choice of $b=0.65$, is discussed in Appendix~\ref{apdx:BIvalidity}.)

\begin{table}[ht]
\renewcommand{\arraystretch}{1.5}
\caption[short desc.]{Quantities related to temporally-bound, indirectly-driven neurons during initial network oscillations characterized by a Binding Index threshold of $b = 0.65$.}
\label{Tbl:BIdata065}
\begin{center}
\begin{tabular}{l|c|}
\multicolumn{2}{l}{Number of bound,  indirectly-driven neurons present:} \\ \hline
Odor 1 & 35\\
Odor 2 & 31\\
Odor 3 & 33\\ \hline
\multicolumn{2}{c}{ }\\
\multicolumn{2}{l}{Number of shared bound,  indirectly-driven neurons between an odor pair:} \\ \hline
Odor 1 and Odor 2 & 15\\
Odor 1 and Odor 3 & 15\\
Odor 2 and Odor 3 & 15\\ \hline
\multicolumn{2}{c}{ }\\
\multicolumn{2}{l}{Percentage of unique bound,  indirectly-driven neurons for each odor in a pair-wise comparison:} \\ \hline
Odor 1 and Odor 2 &57.1\% for Odor 1; \quad 51.6\% for Odor 2\\ 
Odor 1 and Odor 3 & 57.1\% for Odor 1; \quad 54.5\% for Odor 3\\ 
Odor 2 and Odor 3 & 51.6\% for Odor 2; \quad 54.5\% for Odor 3\\ \hline 
\multicolumn{2}{c}{ }\\
\multicolumn{2}{l}{Neuron numbers for bound,  indirectly-driven neurons present in only one of the three odors:} \\ \hline
Odor 1 & \#33, 41, 43, 45 and 59\\
Odor 2& \#73\\
Odor 3& \#2, 3 and 14\\ \hline
\end{tabular}
\end{center}
\end{table}

\begin{figure*}[h!]
\begin{centering}
\includegraphics[width = \textwidth]{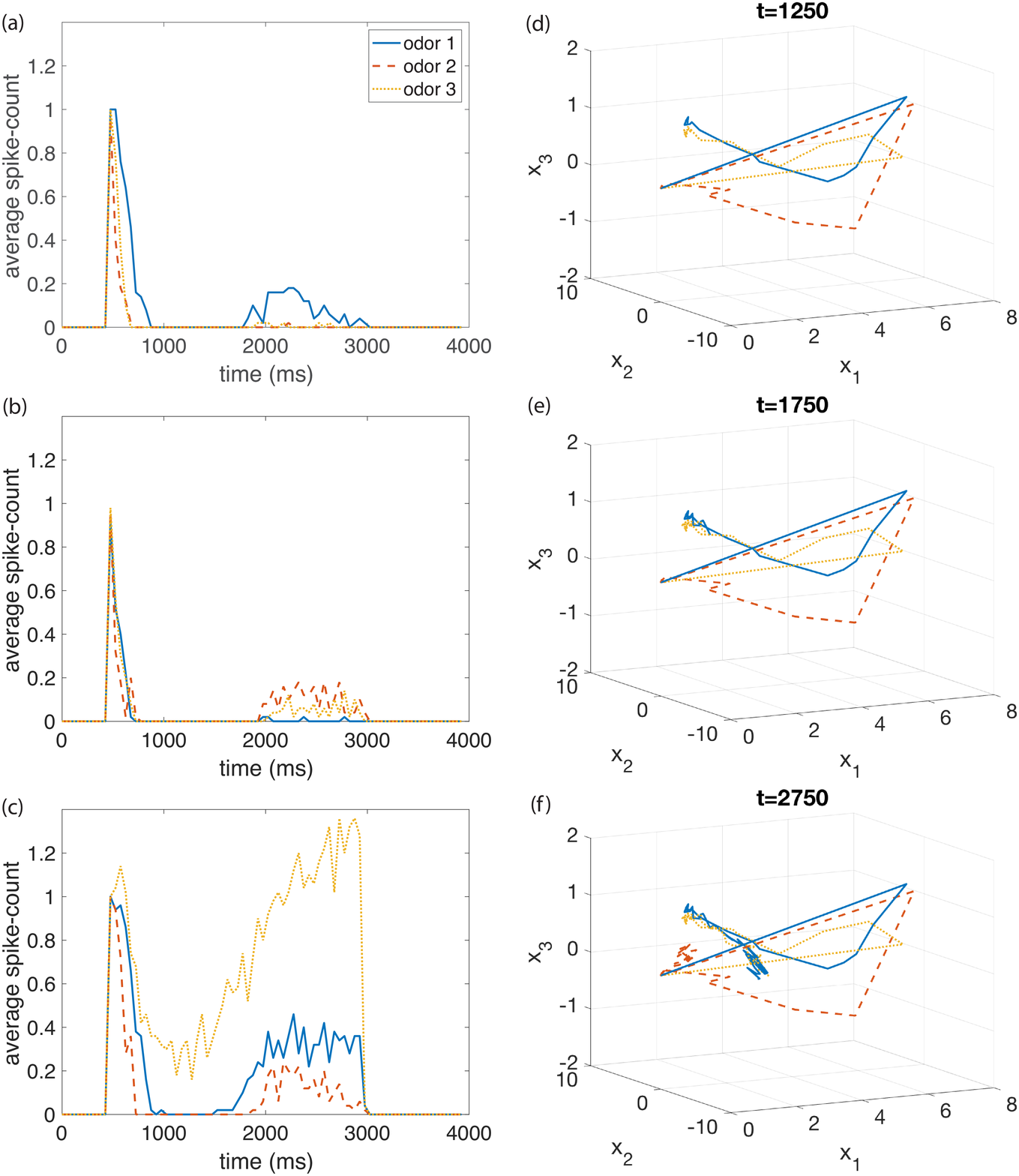}
\caption{Network dynamics corresponding to the presentation of three different odors, generated by stimulating a different 1/3 of the neurons in the network.  (a,b,c): The averaged spike count of 50 simulations binned into 50~ms bins for three different excitatory neurons when presented with the different odors.  (d,e,f): The first 3 components of PCA decomposition for the presentation of the three different odors for the length of time in milliseconds since odor onset indicated.  Computed using parameters in Table~\ref{Tbl:LargeParameterList}.}
\label{fig:odor_PCA}
\end{centering}
\end{figure*}

At longer time scales, after the initial oscillatory phase, our I\&F model also reproduces the experimental results in ref.~\cite{Mazor:2005eu}, in that our model shows slow neuronal firing-rate patterns that differ among presentations of different odors~\cite{Bazhenov:2001pd,Bazhenov:2001ai,PRC09}. A means to visualize the different network responses to different odors is to plot reduced-dimensional firing-rate trajectories of the network.  First, the single-neuron simulation-averaged spike counts in 50~ms bins, shown in Fig.~\ref{fig:odor_PCA}(a), (b), and (c) for the three different odors also used to produce Fig.~\ref{fig:odor_raster}, already display differences.  These differences are enhanced by considering the PCA decomposition of the high-dimensional firing-rate trajectory for the entire set of excitatory neurons in the network.  Using the binned spike counts of these excitatory model neurons, averaged over 50 simulations, as the excitatory network trajectory, we see in Fig.~\ref{fig:odor_PCA}(d), (e), and (f) that the first three components of the PCA decomposition already discriminate among different odors. These reduced trajectories follow different transients to a stationary point over the first second of odor presentation.  The trajectories then remain near this stationary point, corresponding to the lull in firing, until about 1.75 seconds after the odor onset.  Then, the stationary point begins to drift until at 3 seconds the odor is turned off and the trajectories return directly to zero (not shown).  The fact that these firing-rate trajectories discriminate among odors along their entire time course highlights the contribution of the quiescent period and slow patterning to the odor discrimination at times between about 1000 ms and 4000 ms~\cite{Mazor:2005eu,Bazhenov:2001pd,Bazhenov:2001ai,PRC09}.

\subsection{Idealizations of I\&F Model\label{IDEALIZATIONS}}

We now turn our focus away from the details of the response to, and discrimination of, specific odors, and instead focus on the progression of the dynamics through three stages of behavior: 20~Hz oscillations, followed by quiescence, and then followed by slow patterning.  We are interested in determining if there is a simple, robust structure underlying this progression, drawing from the observation that the above three stages appear robust under the presentations of different odors and their different realizations.  Here, we first test the robustness of this neuronal activity by making the I\&F model progressively more idealized through the removal of the structure in the network drive and architecture.  We find that these idealizations still maintain the three-stage dynamical scenario in the I\&F model network.    Thus, we are led to believe that there exists a robust underlying network mechanism responsible for this scenario, and that a robust bifurcation structure can be employed to describe this mechanism.

\begin{figure}[h]
\centering
\includegraphics[width=\textwidth]{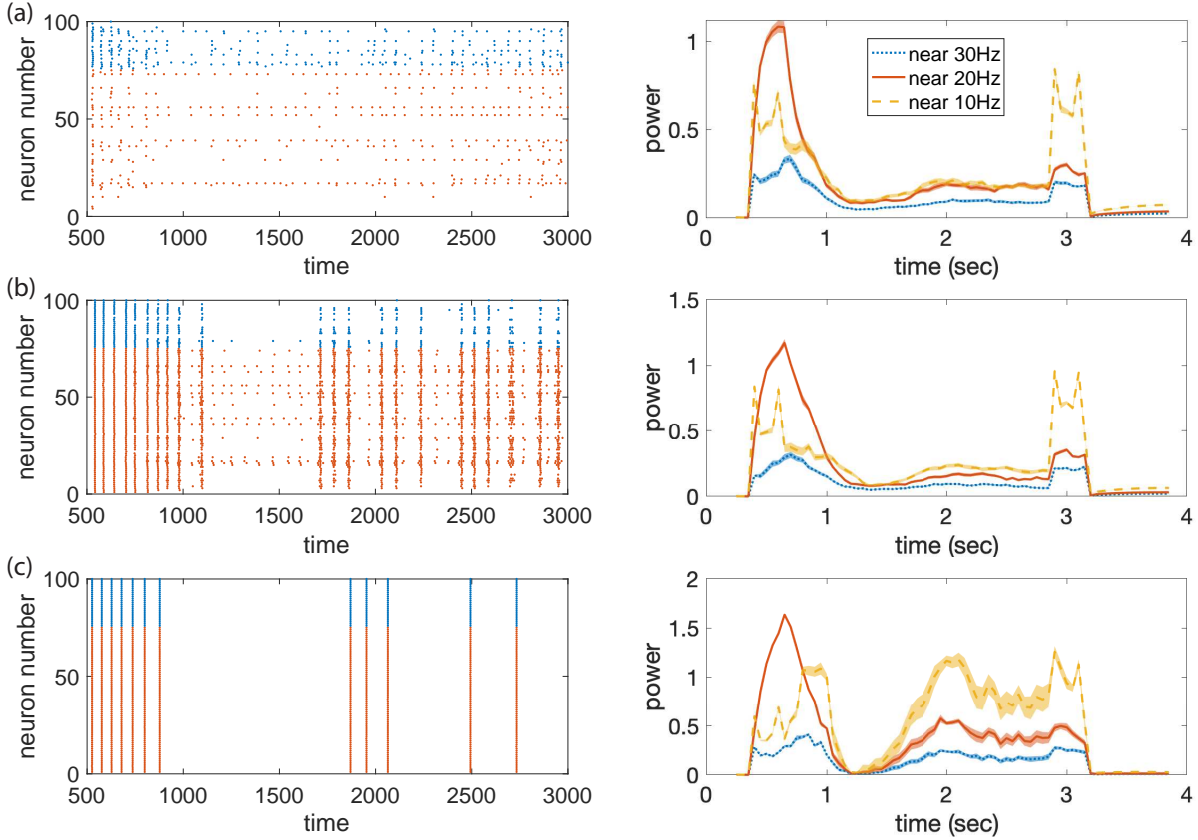}
\caption{
Raster plots (first column) and the total power near 10, 20, and 30~Hz for a moving time-window length of 300~ms over the LFP (second column), displaying the three stages of network behavior for various types of network architecture, receptor locations, stimulus structure, and stimulus-drive properties.  The shaded region indicates the error of the presented sample mean.   (a) Sparse excitatory and inhibitory network connections.  All neurons driven equally by the odor.  (b) Sparse excitatory and inhibitory network connections and more mean-driven input to all neurons.  (c) An all-to-all connected network.  Computed using parameters in Table~\ref{Tbl:RasterParameterList}.} 
\label{fig:rasterplot_transitions}
\end{figure}

Raster plots in the left column of Fig.~\ref{fig:rasterplot_transitions} show network activity for different variants of our I\&F model network. In each case, we consider an increasingly idealized version of the model and bring its structure closer to what can be coarse grained to the FR model, whose dynamics are discussed below in Section~\ref{BIFURCATIONS}. 
In contrast to the case shown in Figs.~\ref{fig:raster_fire_cond} and \ref{fig:PSD}, in which the network is sparse and the stimulus is presented to a subset of the neurons, in the cases shown in Fig.~\ref{fig:rasterplot_transitions}, all neurons receive statistically equivalent, ``white odor" input.  (The reason we call this odor ``white''  is because it contains all possible odors in  the same way as white noise contains all frequencies.)  Presenting a white odor eliminates the question of discrimination among odors, and allows the entire network to participate in each of the initial 20~Hz synchronous firing events.  As listed in Table~\ref{Tbl:RasterParameterList}, the network in   Fig.~\ref{fig:rasterplot_transitions}(a) is more \emph{fluctuation-driven} ($\nu=6$ kHz) than the network in  Fig.~\ref{fig:rasterplot_transitions}(b), which is more \emph{mean-driven} ($\nu=12$ kHz),  while both are still sparse with the same connectivity architecture.   In other words, the (uniform Poisson) afferent firing rate to the neurons in the latter network is increased as compared to that in the former network, whereas the strengths of the synapses, $f_E^{\odor}$ or $f_I^{\odor}$, transmitting the stimulus from the olfactory receptor neurons in the form of the external drive to the AL neurons, are decreased, so that the products of the rates and strengths remain constant.   (In this way, the statistical effect of the drive becomes closer to that of the constant drive employed in the FR model.)
We see that, with each of these two idealizations, namely, dropping the inhomogeneity and reducing the fluctuations in the drive, the regularity of the firing events increases and a larger portion of the network participates in every oscillation cycle.  The network in Fig.~\ref{fig:rasterplot_transitions}(c) is all-to-all connected and mean-driven, and so the most idealized.   It produces the clearest oscillations, with no neurons firing outside of the synchronous events, and will be used for coarse-graining to the FR model. 

The discussion in the previous paragraph describes the sequence of four idealizations of both the network architecture and the form of the odor drive, which are reflected in the  network  dynamics depicted in Figs.~\ref{fig:raster_fire_cond}, \ref{fig:PSD}, and~\ref{fig:rasterplot_transitions}:  sparse connectivity and ORN drive stimulating a specific subset of neurons; sparse connectivity and equally strong (white-odor), fluctuation-driven input to all neurons, sparse connectivity and white-odor, mean-driven input to all neurons; and all-to-all connectivity and white-odor, mean-driven input to all neurons. 
Despite the clear differences in network dynamics among these four types of networks with increasingly idealized properties, as displayed in Fig.~\ref{fig:raster_fire_cond} and the left column of~\ref{fig:rasterplot_transitions}, the three prominent stages of network dynamics remain, as illustrated in the corresponding moving-window-averaged PSD plots, shown in the right columns of Figs.~\ref{fig:PSD}  and~\ref{fig:rasterplot_transitions}. In all the four cases shown in these figures, the power in the frequency interval centered at 20~Hz is strongest during the first 0.5 seconds of odor presentation.
The second stage of suppressed activity is seen in all cases beginning around 1 second into the simulation.  By 2 to 3 seconds, the third stage of activity, characterized by slower oscillations, emerges. During this stage, the moving-window-averaged PSD plots indicate a shift towards power in the frequency interval centered at 10~Hz,
 with the average values of the PSD centered at 20~Hz being lower than those immediately after odor onset. 

The results obtained in this section clearly indicate that the three-stage dynamical scenario consisting of $\sim$ 20~Hz oscillations, quiescence, and slower oscillations or asynchrony represents a robust feature of the AL network dynamics and persists under a sequence of increasing idealizations.    As this scenario is also present in the results of detailed simulations using Hodgkin-Huxley type models~\cite{Bazhenov:2001pd,Bazhenov:2001ai,PRC09,PRC2011}, we conclude that  it is generated by an underlying robust \emph{network} mechanism.    We conjecture this mechanism to be yet more basic, and stem solely from the interaction among the fast excitation and fast and slow inhibition in the AL network.   
Therefore, in the next section, we move from our most idealized, all-to-all connected I\&F network model, driven by a ``white odor" stimulus, to a minimalistic FR model that only considers this interaction in the context of two homogeneous neuronal populations, excitatory and inhibitory.   Using this FR model, we are able to describe the underlying bifurcation structure associated with the robust dynamical scenario established in this section.

\begin{table}[h]
\renewcommand{\arraystretch}{1.5}
\caption[Parameter Values for Raster Picture]{Parameters used for raster plots in Fig.~\ref{fig:rasterplot_transitions}.  In all panels, all neurons receive the odor input at the given rate, and no background stimulus.
 The same network structure is utilized for both the slow and fast inhibitory connections; the corresponding synaptic receptors are colocalized.}
\label{Tbl:RasterParameterList}
\begin{center}
\begin{tabular}{c|c|c|c|}
\multicolumn{1}{c}{} & \multicolumn{3}{|c|}{{\bf Raster Plot Panel}}\\ \hline
\multicolumn{1}{c|}{{\bf Parameter}} & a & b & c  \\ \hline
$N_E$ 									& 75 			 				& 75 				& 75 				\\
$N_I$ 								& 25							& 25				& 25				\\
$\nu$ 						&6\text{ kHz }		&12\text{ kHz }	&12\text{ kHz }	\\
$f_E\nu$						& 13.8 					& 13.8			& 15				\\
$f_I\nu$						& 11.4					& 10.32				& 12				\\

$\sigma_E$ 					& 1\text{ ms }				& 1\text{ ms }		& 1\text{ ms }		\\
$\sigma_F$ 					& 4\text{ ms }				& 4\text{ ms }		& 4\text{ ms }		\\
$\rho_S$ 						& 500\text{ ms }			& 500\text{ ms }	& 500\text{ ms }	\\
$\sigma_S$ 					& 600\text{ ms } 	 		& 600\text{ ms } 	& 600\text{ ms } 	\\
${S^E_E/N_E}$ 					& 0.06					& 0.1				& 0.1				\\
${S^E_I/N_E}$ 						& 0.2 		 			& 0.3 		 	& 0.3				\\
${S^F_E/N_I}$ 						& 1.5 		 	 		 & 0.4 		 	& 0.2 		 	\\
${S^F_I/N_I}$ 						& 0.35		 		 	& 2.0		 		& 1.0		 		\\
${S^S_E/N_I}$ 						& 3.0						& 1.2			& 0.07			\\
${S^S_I/N_I}$ 						& 0.7					& 6.0				& 0.35			\\
$p^E_E$ 							& 0.1			& 0.1				& 1.0				\\
$p^E_I$ 							& 0.1			& 0.1				& 1.0				\\
$p^F_E$ 						   	& 0.15			& 0.15			& 1.0				\\ 
$p^F_I$ 							& 0.25				& 0.25			& 1.0				\\
$p^S_E$ 						 	&0.15 			& 0.15		& 1.0					\\ 
$p^S_I$ 							&0.25 		 		& 0.25		& 1.0					\\
  \hline
\end{tabular}
\end{center}
\end{table}

\subsection{Bifurcation Mechanism\label{BIFURCATIONS}}

In this section, we describe how the three dynamical stages of time evolution exhibited by the results of numerical simulations displayed in Fig.~\ref{fig:raster_fire_cond} --- oscillations followed by quiescence, followed by slower oscillations --- can be explained by a bifurcation structure that follows the amount of slow inhibitory conductance present in the FR model in Eqs.~\eqref{4Dsystem} and \eqref{firing_rates}.   We have justified the relevance of this FR model as a coarse-grained version of the I\&F model by the gradual idealization of cases in Figs.~\ref{fig:raster_fire_cond}, \ref{fig:PSD}, and~\ref{fig:rasterplot_transitions}, where we show that the three stages of the dynamics persist as the I\&F network model transitions through the parameter regimes used in the heuristic derivation of the FR model. By comparing the dynamics of  the effective fast conductances, $g^E$ and $g^F$, for fixed values of the effective slow conductance, $g^S$, to those of the intact FR model in Eqs.~\eqref{4Dsystem} and \eqref{firing_rates}, where all three effective conductances are time-dependent,  we show in this section that the underlying bifurcation structure for these three stages is a slow passage through a SNIC bifurcation~\cite{strogatz:2000,Izhikevich:2007p75,ErmentroutKopell1986}, with the bifurcation parameter $g^S$. 

\begin{figure}[ht]
\centering
\includegraphics[width = 0.5\textwidth]{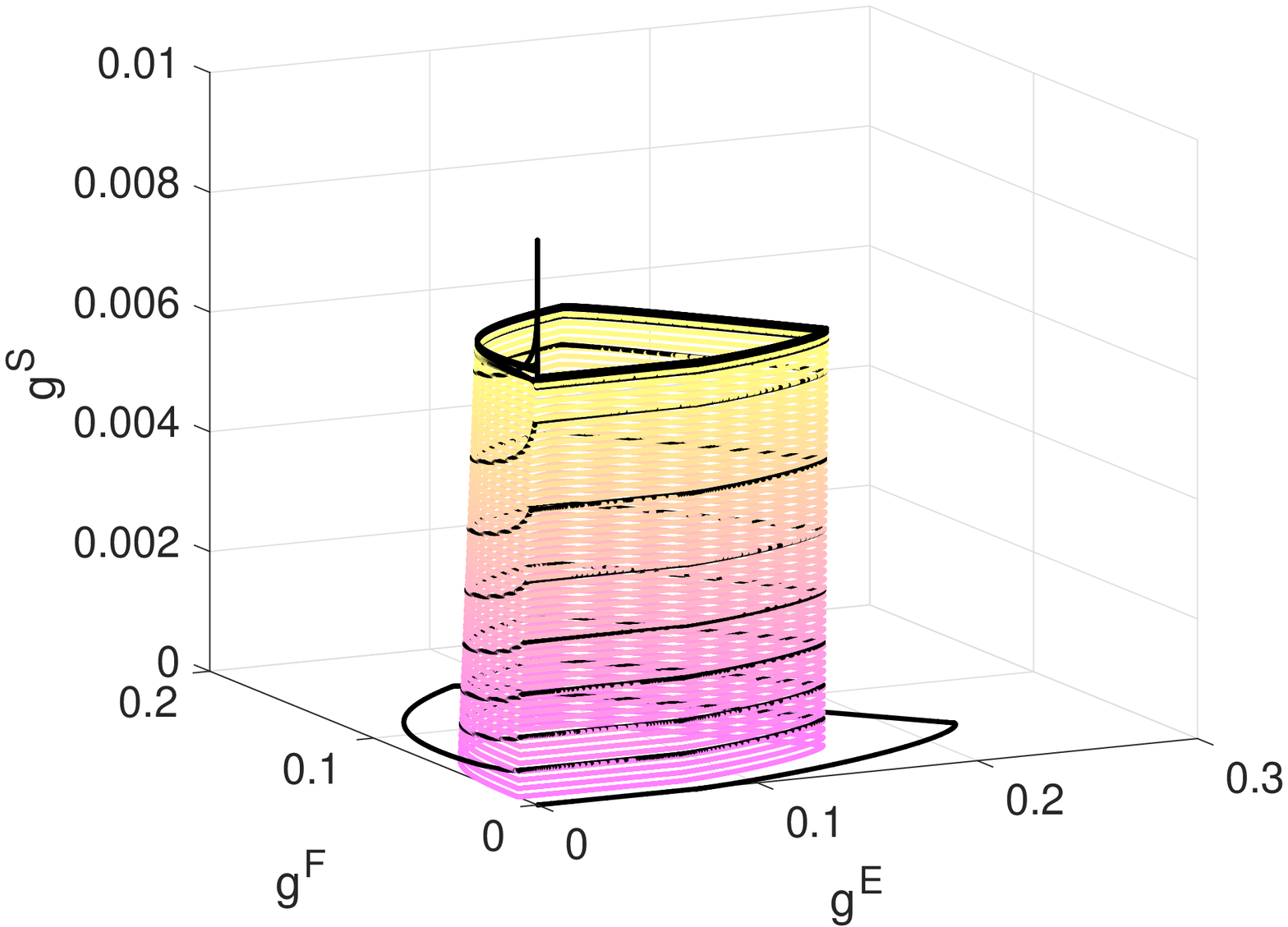}
 \caption{The colored tube depicts a collection of limit cycles in the $g^E$-$g^F$-plane of the nonlinear system in Eqs.~\eqref{4Dsystem} and \eqref{firing_rates} with $g^S$ held constant, ranging from 0 to 0.07.  The solid black line represents the full evolution of the system given by Eqs.~\eqref{4Dsystem} and \eqref{firing_rates}. 
 Parameters are listed in Table~\ref{Tbl:LargeParameterList}. }
\label{fig:gSvaries}
\end{figure}

Oscillations in the FR model in Eqs.~\eqref{4Dsystem} and \eqref{firing_rates} appear to result from the presence of a modulated limit-cycle-like object in the $g^E$-$g^F$ variables.  An attracting limit cycle can indeed be found numerically by holding the slow effective conductance $g^S$ constant (thus eliminating the need for the auxiliary variable $h$), and evolving the remaining two-dimensional system. We computed such limit cycles for fixed values of $g^S$ lying in an interval, and they are shown in Fig.~\ref{fig:gSvaries} as the colored trajectories. These limit-cycle trajectories form a cylindrical object which exists until the bifurcation point in $g^S$ is reached, and then the two-dimensional dynamics change to approaching a stable fixed point.  In Fig.~\ref{fig:gSvaries}, overlaid on this cylindrical object is a trajectory of the intact FR model given by Eqs.~\eqref{4Dsystem} and \eqref{firing_rates} of the same type as the trajectory whose dynamics are shown in Fig.~\ref{fig:raster_fire_cond}(c).  As the slow effective conductance variable $g^S$ increases after the stimulus onset, this trajectory moves up along the set of limit cycles until it passes the bifurcation point.  All the firing is suppressed while this trajectory remains near the line $g^E=g^F=0$.  At this time, the slow effective conductance variable, $g^S$, starts to decay, and this decay moves the system back towards and past the bifurcation point.  The trajectory settles near a limit cycle whose location is controlled by small changes in the value of $g^S$.

\begin{figure}[h!]
\centering
\includegraphics[width = 0.45\textwidth]{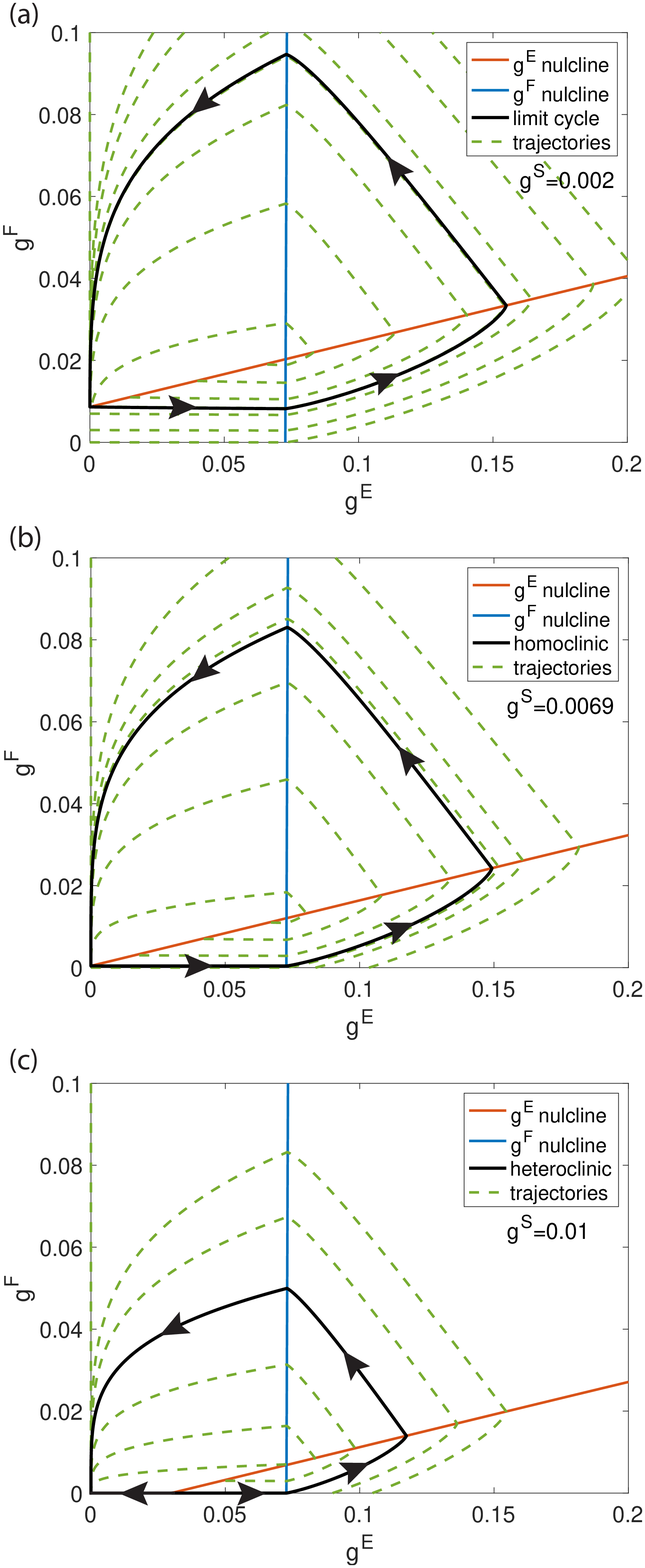}
   \caption{For the nonlinear system in Eqs.~\eqref{4Dsystem} and \eqref{firing_rates} with $g^S$ held constant, the phase-portraits are shown for three values of constant $g^S$. (a) Before the bifurcation point, (b) at the bifurcation point and (c) after the bifurcation point. The limit cycle in panel (a) has turned into a pair of heteroclinic orbits between the saddle and the sink as shown in panel (c).  Parameters in Table~\ref{Tbl:LargeParameterList}.  } 
\label{fig:bifurcation}
\end{figure}

We further investigate the bifurcation with the slow effective conductance variable, $g^S$, held constant, which reduces the four-dimensional FR model in Eqs.~\eqref{4Dsystem} and \eqref{firing_rates} to a two-dimensional system for the fast effective conductances, $g^E$ and $g^F$, with the slow effective conductance, $g^S$, as a parameter.   Different dynamical regimes of this two-dimensional system are summarized in the phase portraits shown in Fig.~\ref{fig:bifurcation}.

Figure~\ref{fig:bifurcation}(a) shows the attracting limit cycle surrounding a source, with no other fixed point in the first quadrant of the $g^E$-$g^F$-plane of the fast effective conductances. (Note that the first quadrant in the $g^E$-$g^F$-plane minus the source can easily be shown to be a trapping region, within which at least one limit cycle must exist by the Poincar\'e-Bendixson theorem.)  As the slow effective conductance, $g^S$, increases, this limit cycle approaches both the $g^E$ and $g^F$ axes.  As shown in Fig.~\ref{fig:bifurcation}(b), the limit cycle first reaches these axes at the origin, and its dynamics stop there.   The origin becomes a degenerate saddle point, with a homoclinic orbit connecting this point to itself, which replaces the limit cycle.  When $g^S$ is increased further, we see in Fig.~\ref{fig:bifurcation}(c) that this degenerate saddle point becomes a true saddle, moves away from the origin along the positive $g^E$-axis, and the origin becomes a sink.  A pair of heteroclinic orbits connect the newly-created saddle and sink, lying close to where pieces of the limit cycle existed before the bifurcation.   Thus, we see that the reduced, two-dimensional FR model undergoes a SNIC bifurcation, with increasing frozen effective slow conductance, $g^S$, as the bifurcation parameter. 

In Appendix \ref{apdx:linearization}, we verify the existence of a unique limit cycle in $g^E$-$g^F$-plane for the linearized version of Eq.~\eqref{4Dsystem}, using an alternative, semi-analytical, approach.

\begin{figure}[h]
\centering
\includegraphics[width = 0.5\textwidth]{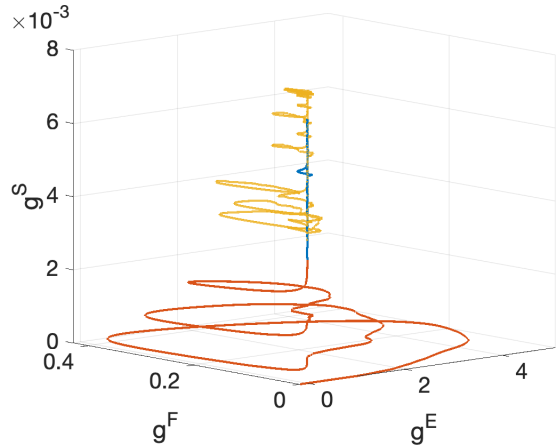}
 \caption{
 Similarly to Fig.~\ref{fig:gSvaries}, we plot the time evolution of 
  the network-averaged effective conductances $\bar{g}^E,~ \bar{g}^F$, and $\bar{g}^S$ given by Eqs.~$\eqref{eq:Network_Avg_Cond}$ for the $N=100$ neurons in the I\&F model, simulated via Eqs.~\eqref{voltage} and~\eqref{conductance}.  Parameters in Table~\ref{Tbl:LargeParameterList}. Red: initial oscillations after odor onset, 500ms to 700ms, Blue: quiescent period, 700ms to 1900ms, Yellow: slow oscillations, after 1900ms.}
\label{fig:networkbif}
\end{figure}

The above-described bifurcation structure explains the transition of the I\&F model dynamics through the three stages of evolution described in Section~\ref{INSECT_OLFACTION}. To demonstrate this claim, in the I\&F network, we compute the trajectory of network-averaged effective conductances based on definitions analogous to those given by Eqs.~\eqref{eqn:substitutions}, and show that it closely resembles the corresponding dynamical trajectory of the FR network. Specifically, we consider the quantities 
\begin{subequations}\label{eq:Network_Avg_Cond}
\begin{align}
\bar{g}^E &= \frac{1}{N} \sum_{i=1}^N \frac{g_i^E - f_i\nu - f_i^\odor\nu^\odor}{S_i^E p_i^E} ,\\
\bar{g}^F &= \frac{1}{N} \sum_{i=1}^N \frac{g_i^F }{S_i^F p_i^F}, \\
\bar{g}^S &= \frac{1}{N} \sum_{i=1}^N \frac{g_i^S }{S_i^S p_i^S},
\end{align} 
\end{subequations}
where the connection probabilities, $p^E_i$, $p^F_i$, and $p^S_i$, and coupling strengths, $S^E_i$, $S^F_i$, and $S^S_i$,  are described in the paragraphs below Eqs.~\eqref{conductance}.  (They are also the same as in assumption~\ref{assum2} in Appendix~\ref{apdx:firingratemodel}.) The trajectory of these network-averaged conductance variables, generated using the I\&F model dynamics shown in Fig.~\ref{fig:raster_fire_cond}(a) (with parameters given in Table~\ref{Tbl:LargeParameterList}),  is plotted in Fig.~\ref{fig:networkbif}.  This trajectory highlights how the three-stage dynamical scenario in Fig.~\ref{fig:raster_fire_cond}(a) can be viewed as underpinned by a slow passage through the SNIC bifurcation described above. 

As we see in Fig.~\ref{fig:networkbif}, initially, the value of the slow inhibitory network-averaged conductance, $\bar{g}^S$, is near zero, while the values of the excitatory and fast inhibitory network-averaged conductances, $\bar{g}^E$ and $\bar{g}^F$, respectively, change in a cyclical manner. This behavior roughly corresponds to the initial oscillatory dynamics of the FR model, during which the slow inhibitory effective conductance has a slow rise time and the system operates in a regime corresponding to the presence of the limit cycle in the fast-variable system with frozen slow effective conductance, $g^S$. Then, the slow network-averaged conductance rises and the activity of the fast network-averaged conductances decreases as the trajectory passes near the bifurcation point corresponding to the bifurcation point in the FR model and the network firing is greatly reduced, similarly to the corresponding FR model dynamics.  Subsequently, the slow network-averaged inhibitory conductance $\bar{g}^S$ decreases and the activity of the fast network-averaged conductances $\bar{g}^E$ and $\bar{g}^F$ again increases, but the values  of the later two remain below their initial activity levels.  We also see that the trajectory in Fig.~\ref{fig:networkbif} is subject to fluctuations present in the simulations, which are absent from the FR model.

Note that in the two-dimensional model in which the slow effective conductance, $g^S$, is held frozen as a parameter, increasing the constant $g^S$ value has the same effect as decreasing the values of the external drives $f_E\nu$ and $f_I\nu$.  The suppressed firing state is equivalent to the system receiving little or no external drive, as one might expect.

%
\section{Discussion}\label{DISCUSSIONS}

In summary, for the early olfactory system dynamics in the locust, we have identified a plausible basic underlying network mechanism, which is the interaction among fast excitation and fast and slow inhibition. 
We have found a highly idealized description of this interaction consisting of a four-component, slow--fast FR model, which reproduces the neuronal network dynamics corresponding to odor detection as a slow passage through a SNIC bifurcation.   The geometric framework on which this bifurcation takes place consists of a cylinder of modulated fast limit cycles, which, after the bifurcation, turn into attracting modulated equilibria.   The modulation consists of a drift along the slow conductances. By tracing, during the simulations of the least-idealized I\&F network discussed in Section~\ref{INSECT_OLFACTION}, the collective, network-wide conductance variables analogous to the effective conductances we use in the FR model, we provide in Fig.~\ref{fig:networkbif} strong numerical evidence for the identified network mechanism and the associated bifurcation structure. 

As stated in the Introduction, we are primarily interested in the basic mechanism governing the dynamics of early olfactory pathways.   Specifically, we would like to understand whether, underneath the structural differences of early olfactory systems among species, there lie in fact some similarities that can be uncovered by model simulation, reduction, and theoretical analysis.  
The details of the network connectivity architecture in the antennal lobe (AL) of the frequently-studied insect species mentioned in the Introduction increase in complexity from the locust~\cite{MacLeod:1996th} through the fruit fly~\cite{Ng:2002ij,Tanaka:2009uq} and sphynx moth~\cite{Heinbockel:1998ys} to the honey bee~\cite{Carcaud2016,Barbara:2005rw}.   For example, glomeruli, densely connected subnetworks of neurons that receive input from one or a very small number of ORN types, feature much more prominently in these other insects, and so their existence and role cannot effectively be neglected in the AL modeling  as it has been for the locust  \cite{Gascuel:1991oz,Kay:2006ea}. Their inclusion brings additional structure to the model network architecture, perhaps one resembling ``small world" connectivity~\cite{Watts:1998aa}, with the glomeruli modeled by densely connected clusters that are loosely connected to one-another~\cite{Rangan:2012kx,Rangan2016}.  Yet further degrees of network complexity arise from sources such as dual inhibitory networks (GABA and glutamatergic) and segregated antennal nerve tracts in the honey bee~\cite{Carcaud2016} and cockroach~\cite{Hidehiro2017}, or separate pathways for pheromone detection in the sphinx moth~\cite{Lei11108}.   Correspondingly, the AL network oscillations seem the least prominent in the fruit fly~\cite{Turner2008,Tanaka:2009uq} and are more prominents in the locust~\cite{Laurent:1996bh}, honey bee~\cite{Stopfer:1997ve}, and  sphinx moth~\cite{Heinbockel:1998ys}.    These differences in details notwithstanding, 
developing new models for different insects and coarse graining them (or some existing ones~\cite{Rangan:2012kx,Rangan2016}), should reveal a hierarchy of idealizations, in which the common structural underpinnings should, as we hope, be easy to discern, hypothesize, and single out for their robustness.  

On a yet broader scale, again as mentioned in the Introduction, early olfaction is similar even across different phyla~\cite{Hildebrand:1997kx}.  For example, glomeruli~\cite{Hildebrand:1997kx,Kay:2006ea} and collective network oscillations are present in the dynamics within early olfaction pathways in vertebrates and molluscs~\cite{Kashiwadani:1999,RojasLbano2008OlfactorySG,Kay:2009bh,KAY2015} and so it would be of interest to explore further parallels between the mechanisms and structures we describe and propose here and their possible analogs present in the olfactory systems in other phyla.    Idealized modeling of olfactory systems across animal phyla, again followed by coarse graining, may thus enable us to formulate  hypotheses of possible yet more general common structural underpinnings and plausible physiological mechanisms of early olfaction that would supplement the wealth of common features brought out by the experimental data.    The interplay among the fast and slow scales in the dynamics, as well as the onset and extinction of synchronous oscillations, point to a possible underlying idealized bifurcation structure consisting of slowly-modulated periodic solutions and fixed points, that is, a structure closely related to the one described in this paper.  Investigating its possible presence will be the task of future work.

\appendix

\section*{Appendix}
 \label{APPENDIX}

\section{Firing-Rate Model Derivation} \label{apdx:firingratemodel}

In this appendix, we present the details of obtaining the FR model, Eqs.~\eqref{4Dsystem} and \eqref{firing_rates} and the inequality~\eqref{boundarycondition2again} presented in Section~\ref{sec:firingratemodel},
 starting from the I\&F model in Eqs.~\eqref{voltage} and \eqref{conductance} driven by a ``white-odor" stimulus.   We, in fact, derive the FR model for a slightly more general connectivity architecture than the all-to-all coupled network, discussed in Section~\ref{sec:firingratemodel}, namely, unstructured, uniform random connectivity as assumed for the most general version of the I\&F model in Section~\ref{IFmodel}.

We heuristically derive the FR model by employing  the following assumptions: 
\renewcommand{\labelenumi}{(\roman{enumi})}
\renewcommand\theenumi\labelenumi
\begin{enumerate}
\item \label{assum5} We assume that the excitatory and fast inhibitory conductances rise instantaneously compared to the slow inhibitory conductance.
\item \label{assum1} We consider the mean-driven regime, in which both network-generated and external-drive spikes are small and arrive at high rates.  Sums over incoming spike trains in Eqs.~(\ref{conductance}) are thus replaced by continuous functions representing the incoming firing rates, since they produce the same conductance dynamics. 
\item \label{assum2} Individual synaptic connections are replaced by the corresponding connection probabilities, $p^E_i$, $p^F_i$, and $p^S_i$, where each probability takes one of the corresponding two values, $p^P_Q$, $P=E$, $F$, and $S$, $Q=E$ or $I$, depending on the $i^{\th}$ neuron's type, as described in the second paragraph below Eqs.~\eqref{conductance}. In this way,  firing rates are scaled by the average number of synaptic connections.  Together with the coupling strengths, $S^E_i$, $S^F_i$, and $S^S_i$, where each coupling strength likewise takes one of the corresponding two values, $S^P_Q$, $P=E$, $F$, and $S$, $Q=E$ or $I$, the network-drive terms in Eqs.~(\ref{conductance}) become $S^E_ip^E_im_E(t)$, $S^F_ip^F_im_I(t)$, and $S^S_ip^S_im_I(t)$, respectively. 
\item  \label{assum4} We assume that the per-neuron firing rates, $m_E(t)$ and $m_I(t)$, vary slowly in comparison to changes in the membrane potential and further assume that the total conductance is high, allowing us to treat the voltage in Eq.~\eqref{voltage} as a constant coefficient differential equation. We can then directly solve for $m_E(t)$ and $m_I(t)$ as the multiplicative inverses of the time for the corresponding voltage solutions to reach threshold from reset. 
\item We neglect the refractory period.
\end{enumerate}
 
Using Assumption~\ref{assum5}, we can adiabatically eliminate the auxiliary variables $h_i^E(t)$ and $h_i^F(t)$ in Eqs.~\eqref{heqn} and \eqref{heqn2}, allowing the incoming spikes to cause instantaneous jumps in the fast conductances, $g_i^E(t)$ and $g_i^F(t)$, respectively.  In other words, we  describe the dynamics of the fast conductance variables themselves using only first-order kinetics in contrast to Eqs.~\eqref{conductance} in the I\&F model which use second-order kinetics. 
We thus replace Eqs.~\eqref{conductance} with the equations
\begin{subeqnarray} \label{conductance_apdx}
\sigma_E \frac{dg_i^E(t)}{dt} &=& -g_i^E(t) +f_{i}^{\odor} \sum_{l} \del(t - \tau^i_l)+ f_{i} \sum_{k} \del(t - \gamma^i_k) \nonum\\
&& + \frac{S^E_{i}}{N_E}\sum_{j \neq i}p^E_{ji}\sum_{\mu}\del(t-t^j_{\mu}), \slabel{gE_instantrise} \\
\sigma_F \frac{dg_i^F(t)}{dt} &=& -g_i^F(t) + \frac{S^F_{i}}{N_I}\sum_{j \neq i}p^F_{ji}\sum_{\mu}\del(t-t^j_{\mu}), \slabel{gF_instantrise}\\
\sigma_S \frac{dg_i^S(t)}{dt} &=& -g_i^S(t) +h^S_i(t), \slabel{gS_apdx}\\
\rho_S \frac{dh_i^S(t)}{dt} &=& -h_i^S(t) + \frac{S^S_{i}}{N_I}\sum_{j \neq i}p^S_{ji}\sum_{\mu}\del(t-t^j_{\mu}), \slabel{hS_apdx} 
 \end{subeqnarray}
where all variables and parameters are defined for Eqs.~\eqref{conductance} in Section~\ref{IFmodel}.   Consequently, we must readjust the values of the excitatory and fast inhibitory conductance decay rates, $\sigE$ and $\sigF$, respectively, so that the conductance responses have the appropriate duration lengths.

The next step is to consider the mean-driven limit of the incoming spikes in the model driven by a ``white odor" stimulus.    To mimic the ``white odor" stimulus, the external spikes from the background and odor are combined to form a single source of external input with Poisson rate $\nu$, and spike strengths $f_E$ and $f_I$ for the excitatory and inhibitory neurons, respectively. We then further assume that the external drive operates in the mean-driven regime, in which each individual spike is small but the spikes arrive at high rates (i.e., $f_i\to 0,~\nu\to\infty$ with $f_i\nu$ held constant), replacing the sums over the incoming spike trains in Eq.~\eqref{gE_instantrise}, $f_{i}^{\odor} \sum_{l} \del(t - \tau^i_l)+ f_{i} \sum_{k} \del(t - \gamma^i_k)$, with their statistical averages, the constant functions $f_i\nu$, where $i = E$ or $I$ depending on whether the $i^\th$ neuron belongs to the excitatory or inhibitory population, respectively.   

Furthermore, as we only consider networks with unstructured, uniform random connectivity, we can assume that probability of a synaptic connection between any pair of neurons of given types will be statistically equivalent to that of any other synaptic connections between any other pair of the same types of neurons. Thus, statistically, we can replace the individual synaptic connection weights, $p^E_{ij}$, $p^F_{ij}$, and $p^S_{ij}$, by the corresponding connection probabilities between the corresponding populations, $p_Q^E$, $p_Q^F$, and $p_Q^S$, respectively.

From the coarse-graining process carried out in the previous two paragraphs, we can conclude that the resulting conductance equations  describe the conductance dynamics of a typical (excitatory or inhibitory) neuron, i.e., one that is statistically equivalent to all other neurons of the same type. Thus, we replace the individual neuronal conductance equations in Eqs.~\eqref{conductance_apdx} by equations for the conductances of the corresponding neuronal populations. 
This replacement eliminates the need for the index $i$  and lets us introduce the subscript notation of $Q$ in its place, where $Q$ is either $E$ or $I$ to represent quantities associated with the typical excitatory or inhibitory neuron, or, equivalently, the corresponding populations, respectively. 

We now apply the same mean-driven limit to the network-generated spikes as we did to the external spikes.  Due to the uniform, unstructured random connectivity of the network, we can assume that each excitatory and inhibitory neuron, respectively, is driven by the average per-neuron firing rate of the appropriate type, $m_E(t)$ and $m_I(t)$.   We scale these by the number of neurons of each type in the network and their synaptic connection probabilities to find the population-averaged firing rates, $N_Ep^E_Qm_E(t)$, $N_Ip^F_Qm_I(t)$, and  $N_Ip^S_Qm_I(t)$, 
arriving at the acetycholine, GABA$_A$, and slow inhibitory receptors, respectively, on the 
postsynaptic excitatory ($Q=E$) and inhibitory ($Q=I$) neurons.    Thus, the sums over the network spikes, $\frac{S^E_{i}}{N_E}\sum_{j \neq i}p^E_{ji}\sum_{\mu}\del(t-t^j_{\mu})$, $\frac{S^F_{i}}{N_I}\sum_{j \neq i}p^F_{ji}\sum_{\mu}\del(t-t^j_{\mu})$, and $ \frac{S^S_{i}}{N_I}\sum_{j \neq i}p^S_{ji}\sum_{\mu}\del(t-t^j_{\mu})$, in Eqs.~\eqref{gE_instantrise}, \eqref{gF_instantrise}, and \eqref{hS_apdx}, are replaced by the average network drive terms $S^E_Qp^E_Qm_E(t)$, $S^F_Qp^F_Qm_I(t)$, and $S^S_Qp^S_Qm_I(t)$, respectively.  

In this way, since each excitatory or inhibitory neuron is statistically equivalent to all other neurons of the same type, the set of $4N$ equations for the neurons' conductances in Eqs.~\eqref{conductance_apdx} can be replaced by the following eight representative equations for the two neuronal populations:
\begin{subequations}\label{8eqnsystem}
\begin{eqnarray}
\sigE \frac{dg_Q^E}{dt} &=& -g_Q^E+ f_Q\nu +S^E_Qp^E_Qm_E(t), \label{gEE_full}\\
\sigF \frac{dg_Q^F}{dt} &=& -g_Q^F+S^F_Qp^F_Qm_I(t), \label{gEF_full}\\
\sigS\frac{dg^S_Q}{dt}&=& -g^S_Q + h_Q, \label{gSE_full}\\
\rho_S\frac{dh_Q}{dt}&=& -h_Q +S^S_Qp^S_Qm_I(t), \label{hE_full}
\end{eqnarray}
\end{subequations}
with $Q=E$ or $I$ for the excitatory or inhibitory population, respectively.

To solve for the firing rates, we treat the voltage equation in Eq.~\eqref{voltage} as a constant-coefficient differential equation; we have one such equation for each of the two populations.  This is justified by assuming that the per-neuron firing rates, $m_E(t)$ and $m_I(t)$, and therefore also the associated conductances, vary slowly compared to the voltages, while also assuming that the conductances are relatively high so that the voltages vary fast. This means that the (conductance-induced) time scale of the voltage is shorter than the shortest conductance time scale (see ref.~\cite{SMSW02}).   For constant conductance values $g^E_Q,~g^F_Q,$ and $g^S_Q$, the first-order linear equation in Eq.~\eqref{voltage} becomes a constant coefficient equation, and takes the form
\beq\begin{aligned}
\tau\frac{dv_Q(t)}{dt} =& -\left(1+g^E_Q+ g^F_Q + g^S_Q\right)v_Q(t) + \left(\vs_R + g^E_Q \vs_E + g^F_Q\vs_F + g^S_Q\vs_S\right),
\end{aligned}\eeq
which has the solution
\beq\begin{aligned} \label{Veqn_solved}
v_Q(t) &= Ce^{-\left(1+g^E_Q+ g^F_Q + g^S_Q\right)t/\tau} + \frac{\vs_R + g^E_Q \vs_E + g^F_Q\vs_F + g^S_Q\vs_S}{1+g^E_Q+ g^F_Q + g^S_Q},
\end{aligned}\eeq
with arbitrary constant of integration, $C$. This constant is determined by the initial condition that the membrane potential starts at its resting potential value, $\vs_R$, just after spiking,
\beq \label{Vatzero}
v_Q(0) = \vs_R, \eeq
and thus,
\beq \label{constantC}
C = \frac{g^E_Q(\vs_R - \vs_E)+g^F_Q(\vs_R -\vs_F)+g^S_Q(\vs_R -\vs_S)}{1+g^E_Q+g^F_Q+ g^S_Q}.
\eeq

At the time $\ts$, the membrane potential in Eq.~\eqref{Veqn_solved} with $C$ in Eq.~\eqref{constantC}, reaches the threshold, 
\beq  \label{Vatspike}
v_Q(\ts) = V_T.\eeq
Therefore, $\ts$ is the amount of time between a neuron's consecutive spikes given the particular values of the conductances. The  reciprocal of $\ts$ is thus the firing rate of the typical neuron in the population, $m_Q$, at those conductance values.  

The membrane potential, $v_Q$, only crosses the firing threshold when the corresponding slaving potential, 
\begin{align}\label{slavingV}
 V_{s,Q} = \frac{ \vs_R+ g^E_Q \vs_E+g^F_Q\vs_F+g^S_Q\vs_S}{1+g^E_Q+g^F_Q+g^S_Q},
 \end{align}
is larger than the spiking threshold, $V_T$, as the membrane potential is always being drawn toward $V_{s,Q}$. 
The condition $ V_{s,Q} > V_T$ can be rewritten as a condition on the effective excitatory conductance $g^E_Q$ such that 
\begin{align}\label{ineq:Vs_VT_gE}
g^E_Q > \frac{V_T - \vs_R}{\vs_E - V_T} + g^F_Q\left(\frac{V_T - \vs_F}{\vs_Q - V_T}\right)+g^S_Q\left(\frac{V_T - \vs_S}{\vs_Q - V_T}\right).
\end{align}
Using Eqs.~\eqref{Veqn_solved} through \eqref{Vatspike}, we can calculate the spike time $\ts$ for the effective conductance region where these conditions hold, arriving at the expression
\begin{align}\label{tspikei_eqn}
\ts =&\frac{\tau}{1+g^E_Q+g^F_Q+g^S_Q} \ln\left[\tfrac{g^E_Q(\vs_E - \vs_R)+g^F_Q(\vs_F - \vs_R)+g^S_Q(\vs_S- \vs_R)}{\left\{\Delta\vs_R+ g^E_Q\Delta \vs_E+g^F_Q\Delta\vs_F+g^S_Q\Delta\vs_S\right\}^+}\right],
\end{align}
where $\Delta\vs_Z$ is defined in Eq.~\eqref{deltaz} and $\{ \cdot \}^+$ is defined as after Eqs.~\eqref{firing_rates}. 

As stated above, the reciprocal of $\ts$ in Eq.~\eqref{tspikei_eqn} is the per-neuron firing rate $m_Q$, given by the equation
\beq\label{meqns_A2AEfastIslowI1}
m_Q(t) = \frac{1+g_Q^E+ g^F_Q+ g^S_Q}{\tau \ln\left[\frac{g_Q^E(\vs_E - \vs_R)+ g^F_Q(\vs_F-\vs_R)+g^S_Q(\vs_S-\vs_R)}{ \left\{\Delta\vs_R + g_Q^E \Delta\vs_E + g^F_Q\Delta\vs_F+g^S_Q\Delta\vs_S\right\}^+ }\right]},
\eeq
where, again, $Q=E$ or $I$.
Note that, in the case when $ V_{s,Q} < V_T$, no spiking occurs and the average time to spike, $\ts$, would be taken as infinite, so that the per-neuron firing rate vanishes.   The firing rate, $m_Q$ in Eq.~\eqref{meqns_A2AEfastIslowI1},  is inserted into the conductance equations in Eqs.~\eqref{8eqnsystem} to close the FR model.

As stated in the previous paragraph, when the denominator inside the logarithm in Eq.~(\ref{meqns_A2AEfastIslowI1}) vanishes, we define the corresponding firing rate, $m_Q$, $Q \in \left\{E,I\right\}$, to also vanish.  Thus, we conclude that the firing rate $m_Q(t)$ is a piecewise-defined, continuous function with $m_Q(t)=0$ when 
\beq\label{boundarycondition1}
\Delta\vs_R + g_Q^E \Delta\vs_E + g^F_Q\Delta\vs_F+g^S_Q\Delta\vs_S \le 0,
\eeq
and is given by Eq.~\eqref{meqns_A2AEfastIslowI1} otherwise.

Using the relations 
\begin{subequations}\label{eqn:substitutions_a}
\begin{eqnarray}
g^E_Q &= &S^E_Qp^E_Qg^E+f_Q\nu, \label{sub_gE} \\
g^F_Q &= &S^F_Qp^F_Qg^F, \label{sub_gF}\\
g^S_Q &= &S^S_Qp^S_Qg^S, \label{sub_gS}\\
h_Q &= &S^S_Qp^S_Qh,  \label{sub_h}
\end{eqnarray}
\end{subequations}
pairs of equations in Eqs.~\eqref{8eqnsystem} for each type of conductance become redundant and the FR system, defined by Eqs.~\eqref{8eqnsystem} and \eqref{meqns_A2AEfastIslowI1}  and the inequalities in Eqs.~\eqref{boundarycondition1}  can be reduced to the four-equation model for the effective conductance variables $g^E, g^F, g^S$, and $h$, given by the conductance equations in Eqs.~\eqref{4Dsystem}   presented in Section~\ref{sec:firingratemodel}, together with the firing-rate equations 
\begin{subequations}\label{firing_rates_a}
\beq\label{meqns_simple_a}
m_Q(t) =\frac{1+S^E_Qp^E_Qg^E+f_Q\nu+S^F_Qp^F_Qg^F+S^S_Qp^S_Qg^S}{\tau \ln\left[M_Q\right]} ,
\eeq
where
\beq
M_Q  = \frac{ \sum_{P\in\{E,F,S\}} S^P_Qp^P_Qg^P(\vs_P - \vs_R)+f_Q\nu(\vs_E - \vs_R)}{\left\{ \Delta\vs_R   +f_Q\nu\Delta\vs_E+\sum_{P\in\{E,F,S\}}S^P_Qp^P_Qg^P\Delta\vs_P   \right\}^+ }.
\eeq
\end{subequations}
The conditions for which the firing rates in Eqs.~\eqref{firing_rates_a} do not vanish become
\beq\label{boundarycondition2again_a}
 \Delta\vs_R   +f_Q\nu\Delta\vs_E+\sum_{P\in\{E,F,S\}}S^P_Qp^P_Qg^P\Delta\vs_P > 0.
 \eeq

For an all-to-all coupled network, we take the connection probabilities $p_Q^E=p_Q^F=p_Q^S=1$, and the above reduced model becomes that given by the equations
in Eqs.~\eqref{4Dsystem} and~\eqref{firing_rates} and the inequalities in Eqs.~\eqref{boundarycondition2again} in Section~\ref{sec:firingratemodel}.

We remark on the above assumption \ref{assum4}, which states that the dynamically-induced membrane-potential time scales are taken to be faster than the conductance time scales.  This is rarely the case.  In particular, the voltage time scale almost never falls below 50\% of the conductance time scale; nevertheless, the frozen-conductance approximation in the voltage equation is quite accurate \cite{SMSW02}.  Therefore, despite this accurate agreement, the derivation of the FR model presented here is not systematic in terms of any small parameter, and thus the FR model cannot be considered as a limit of the I\&F model. While we can expect qualitative agreement between the dynamics of the I\&F model and the FR model, we can only expect to achieve a more quantitative agreement after possibly tuning some of the FR model parameters.

\section{Robustness of the Binding-Index Threshold} \label{apdx:BIvalidity}

In this appendix, we discuss the robustness of our choice for the value of the binding index threshold $b$, defined in Sec~\ref{sec:bidef}.   As mentioned in Sec.~\ref{INSECT_OLFACTION},  the threshold value  $b = 0.65$ we use for the binding index is the same as that used in ref.~\cite{PRC2011}. There, it is shown that this choice of $b$  is suitable to identify a distinct, clearly defined subset of the PNs that are temporally-bound given each odor, and only indirectly driven by the corresponding excited ORNs, which can be used to distinguish among different odors.   This is also true in the results of our simulations, as we have discussed in  Sec.~\ref{INSECT_OLFACTION} and displayed in Fig.~\ref{fig:odor_raster}   and Table~\ref{Tbl:BIdata065}. 

\begin{figure*}[ht]
\centering
         \includegraphics[width = \textwidth]{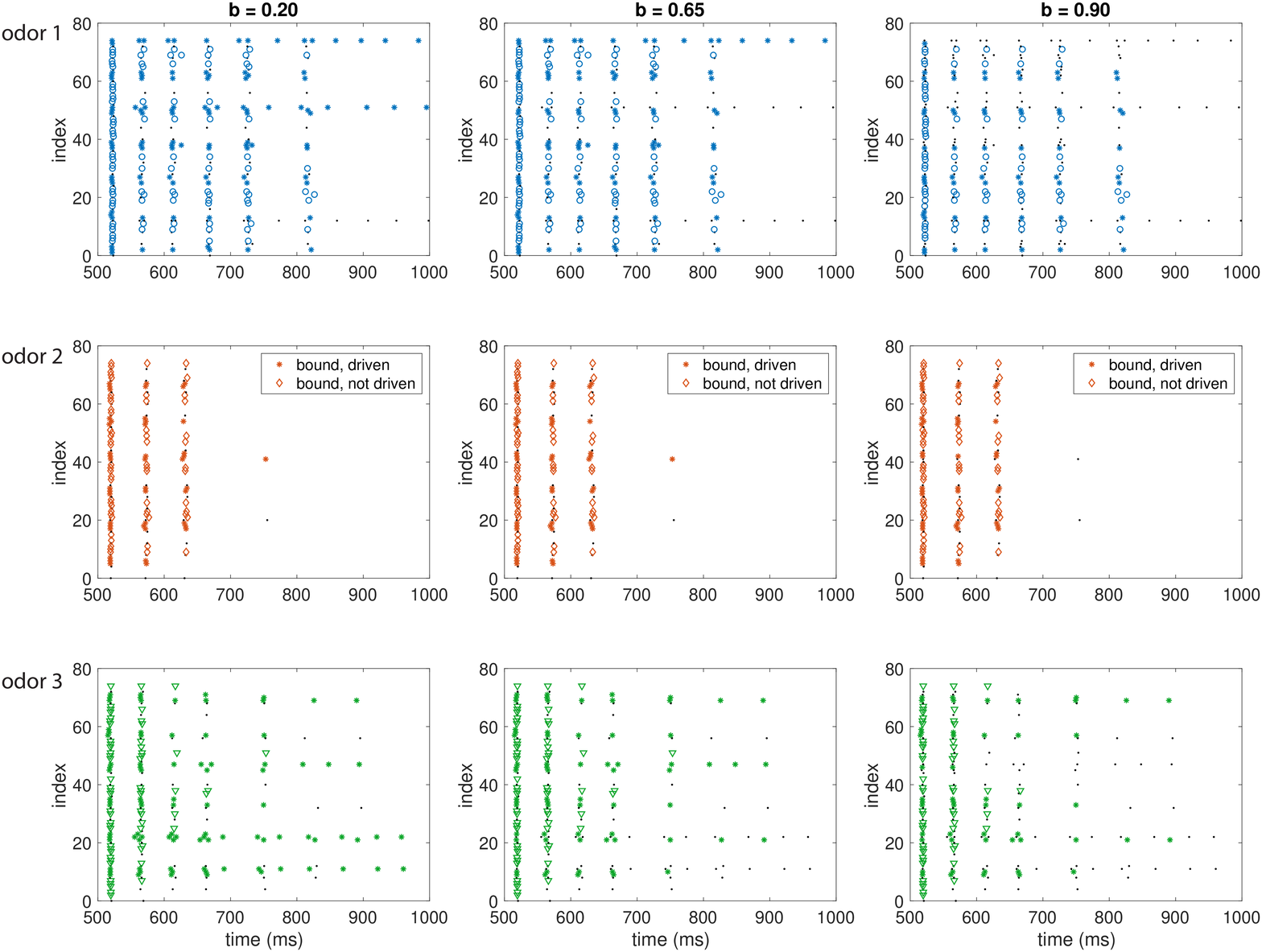}
  	 \caption{Raster plots of the excitatory neurons in a network to which three different odors are presented, generated by stimulating a different 1/3 of the neuron in the network.  Each column corresponds to a different binding index $b$, for which temporally-bound neurons are those with binding index $BI_{i,j,k} \ge b$.  In all panels, black dots represent unbound neurons, stars represent temporally-bound neurons directly driven by the odor, and open symbols represent temporally bound neurons not directly driven by the odor stimulus.  The figure illustrates that many bound neurons are stable with respect to binding index threshold $b$.  Parameters found in Table~\ref{Tbl:LargeParameterList}.}
 \label{fig:bivalidity} 
\end{figure*}

In particular, in Fig.~\ref{fig:bivalidity} we use the results of the same numerical simulations as those used in Fig.~\ref{fig:odor_raster}, but we now analyze them using three different values of the binding-index threshold: $b=0.2$, $b = 0.65$, and $b=0.9$.   (Note that the middle column of Fig.~\ref{fig:bivalidity} is identical to Fig.~\ref{fig:odor_raster}.)    
We see that only a few neurons that are bound at $b = 0.65$ change their bound/unbound property between the values of $b=0.2$ and $b=0.9$, and most of those neurons are driven directly by the odor stimulus arriving from the ORNs. For Odor 1 in the top row of Fig.~\ref{fig:bivalidity} (spiking activity shown in blue), we found that the only indirectly-driven neuron that changes from a bound to unbound classification with the increase in threshold from 0.65 to 0.9 is neuron 69.  For Odor 3 in the bottom row of Fig.~\ref{fig:bivalidity} (spiking activity shown in green), we found that only the indirectly-driven neurons 37 and 51 change their classification. Finally, for Odor 2 in the middle row of Fig.~\ref{fig:bivalidity} (spiking activity shown in red), no indirectly-driven neurons change their bound/unbound classification between threshold values of 0.65 and 0.9. Additionally, when comparing spiking behavior at a threshold of $b = 0.2$ to that at $b = 0.65$ we observe that all indirectly-driven bound, neurons remain bound at both thresholds for all three model odors. Therefore, Table~\ref{Tbl:BIdata065} also provides the data for $b = 0.2$. Overall, the largest change for the odors shown occurs for Odor 1 from $b = 0.65$ to $b = 0.9$ where only 8\% of the PNs (directly- and indirectly-driven combined) change classification from bound to unbound. Thus, the raster plots of bound neurons between the threshold values $b=0.2$ and $b=0.9$ can still clearly discriminate among different odors. 

From the discussion in the previous paragraph, we conclude that when we consider the portion of indirectly-driven, temporally-bound neurons that are uniquely assocaited to a single odor in a pair-wise comparison between two odors for threshold values of $b = 0.65$ and $b = 0.9$, we find that there are overall fewer indirectly-driven, bound neurons at the higher threshold but that a greater than or equal number and a larger percentage of them are uniquely associated to that odor. This holds across all three pair-wise comparisons and is supported by the data provided in Tables~\ref{Tbl:BIdata065} and~\ref{Tbl:BIdata09}. 

We identified this interval from $b = 0.2$ to $b = 0.9$ somewhat arbitrarily to be near, but not exactly equal to the threshold values of $b = 0$ and $b = 1$, respectively. Choosing these extreme values would be prohibitively inclusive at $b = 0$ (neurons in all triplets would be classified temporally-bound) and exclusive at $b = 1$ (only neurons in triplets with identical spiking behavior would be classified as bound).

The discussion in this appendix thus shows that there exists a large interval of threshold values of the binding index, b, for which the odor-discriminability data quantitatively changes little but the ability to discriminate among odors qualitatively changes none at all. 

\begin{table}[ht]
\renewcommand{\arraystretch}{1.5}
\caption[short desc.]{Quantities related to temporally-bound, indirectly-driven neurons during initial network oscillations characterized by a Binding Index threshold of $b = 0.9$.}
\label{Tbl:BIdata09}
\begin{center}
\begin{tabular}{l|c|}
\multicolumn{2}{l}{Number of bound, indirectly-driven neurons present:} \\ \hline
Odor 1 & 30\\
Odor 2 & 31\\
Odor 3 & 31\\ \hline
\multicolumn{2}{c}{ }\\
\multicolumn{2}{l}{Number of shared bound, indirectly-driven neurons between an odor pair:} \\ \hline
Odor 1 and Odor 2 & 13\\
Odor 1 and Odor 3 & 11\\
Odor 2 and Odor 3 & 13\\ \hline
\multicolumn{2}{c}{ }\\
\multicolumn{2}{l}{Percentage of unique bound, indirectly-driven neurons for each odor in a pair-wise comparison:} \\ \hline
Odor 1 and Odor 2 &56.7\% for Odor 1; \quad 58.1\% for Odor 2\\ 
Odor 1 and Odor 3 & 63.3\% for Odor 1; \quad 64.5\% for Odor 3\\ 
Odor 2 and Odor 3 & 58.1\% for Odor 2; \quad 58.1\% for Odor 3\\ \hline 
\multicolumn{2}{c}{ }\\
\multicolumn{2}{l}{Neuron numbers for bound, indirectly-driven neurons present in only one of the three odors:} \\ \hline
Odor 1 & \#33, 41, 43, 45 and 59\\
Odor 2& \#37, 51, 69, and 73\\
Odor 3& \#2, 3, 5, 14, 18, 53, and 65\\ \hline
\end{tabular}
\end{center}
\end{table}

\section{Limit Cycle in the Linearized Firing-Rate Model}   \label{apdx:linearization}

In this appendix, we study a linearization of  the FR model in Eqs.~\eqref{4Dsystem} and~\eqref{firing_rates_a}, with an eye on providing a demonstration that a \emph{unique}, attracting limit cycle exists in the two-dimensional FR model for the fast effective conductances $g^E$ and $g^F$ with frozen slow effective conductance, $g^S$.    Thus, in Appendix~\ref{apdx:linearizedmodel}, we derive this linearization, and in Appendix~\ref{apdx:limitcycle} we semi-numerically verify the existence 
of a unique limit cycle in the resulting planar piecewise-linear system for $g^E$ and $g^F$ with frozen $g^S$.   In particular, we show that we can find the trajectories explicitly in the four effective-conductance regions in which this system is linear, and splice them together using a numerical solution of a transcendental equation at each boundary between two such regions.   We use these spliced trajectories to construct a Poincar\'e map, and demonstrate the existence of the unique limit cycle by finding a unique fixed point of this map.


\subsection{Derivation of the Linearized Model} \label{apdx:linearizedmodel}

In this section, we present the details of linearizing the FR model in Eqs.~\eqref{4Dsystem} and~\eqref{firing_rates_a}, which results in a piecewise-linear model.  

Our procedure is to first note that if a firing rate $m_Q$, $Q =E $ or $I$, vanishes, the corresponding equation(s) in Eqs.~\eqref{4Dsystem} is (are) already linear.  If however, a firing rate is nonzero, we linearize it as a function of the conductances in order to linearize the corresponding equation(s) \eqref{4eqn_gE}, \eqref{4eqn_gF}, or \eqref{4eqn_h}.  This procedure creates a piecewise-linear function for the derivative of each effective conductance variable: $g^E$, $g^F$, and $h$; the firing-rate term does not appear in the differential equation for the slow effective conductance $g^S$, and thus Eq.~\eqref{4eqn_gS} does not take a piecewise form.  A crucial step involves determining the boundaries between the four regions created by the various combinations of parts of the piecewise-defined firing rates: $m_E = m_I = 0$,   $m_E=0$ and $m_I\ne0$, $m_E\ne0$ and $m_I=0$, $m_E, m_I \neq 0$, along which the system switches among the differential equations that govern the trajectory.

When a slaving voltage $V_{s,Q}$  in Eq.~\eqref{slavingV} is subthreshold, $ V_{s,Q} <V_T$, the inequality in Eq.~\eqref{ineq:Vs_VT_gE} is not satisfied, and the corresponding firing rate vanishes, $m_Q(t) = 0$, $Q=E$ or $I$.  It would follow then that each equation in Eqs.~\eqref{4Dsystem} would take a linear form. In the superthreshold case when $ V_{s,Q}  > V_T$, $m_Q(t)$ is nonlinear for each $Q =E$ or $I$.   We linearize about large values of the effective excitatory conductance $g^E$.  (See more discussion at the end of this section.)  In order for our linearization to proceed more systematically, we first expand the firing rates  $m_Q(t)$ about large values of the conductances $g^E_Q$, where $Q =E$ or $I$, and only then express $g^E_Q$ in terms of the effective conductance $g^E$ using formulas in Eqs.~\eqref{eqn:substitutions_a}.

By linearizing the firing rates in the superthreshold case, we can produce a system of piecewise-linear equations for the derivatives of the effective conductances.  We then use the reduced two-dimensional fast model, in which $g^S$ is treated as a frozen parameter, to demonstrate the existence of a limit cycle analytically, save for four root-finding calculations. The details of this analysis are given in Appendix~\ref{apdx:limitcycle}, below. 

To more easily identify the large parameter $g^E_Q$, it is helpful to rearrange the fraction inside of the logarithm in Eq.~\eqref{meqns_A2AEfastIslowI1} as follows:
\beq\begin{aligned}\label{eq:frac1}
&\frac{g^E_Q(\vs_E - \vs_R)+  g^F_Q(\vs_F -\vs_R)+g^S_Q(\vs_S -\vs_R)}{\Delta\vs_R+ g^E_Q \Delta \vs_E+g^F_Q\Delta\vs_F+g^S_Q\Delta\vs_S } \nonumber\\
&\qquad =\frac{A\left[1-\left(Bg^F_Q + Xg^S_Q\right)/g^E_Q\right]}{ 1-\left[(A-1)- \frac{\Delta\vs_F}{\Delta\vs_E}g^F_Q- \frac{\Delta\vs_S}{\Delta\vs_E}g^S_Q\right]/g^E_Q},
\end{aligned}\eeq
with $A$, $B$, and $X$ defined as

\begin{align}\label{ABX}
&A = \frac{\vs_E - \vs_R}{\vs_E - V_T}, \quad B = \frac{\vs_R -\vs_F}{\vs_E - \vs_R}, \quad X = \frac{\vs_R -\vs_S}{\vs_E - \vs_R}. 
\end{align}
The logarithm of Eq.~\eqref{eq:frac1} can be rearranged so that the resulting expression may be expanded for large $g^E_Q$, 

\begin{align}\label{eq:frac2}
\log&\left\{\frac{A\left[1-\left(Bg^F_Q+ Xg^S_Q\right)/g^E_Q\right]}{ 1-\left[(A-1)- \frac{\Delta\vs_F}{\Delta\vs_E}g^F_Q- \frac{\Delta\vs_S}{\Delta\vs_E}g^S_Q\right]/g^E_Q}\right\} = \log A + \log\left[1-\left(Bg^F_Q + Xg^S_Q\right)/g^E_Q\right]  \nonum \\
& \qquad - \log\left\{1-\left[(A-1)- \frac{\Delta\vs_F}{\Delta\vs_E}g^F_Q- \frac{\Delta\vs_S}{\Delta\vs_E}g^S_Q\right]/g^E_Q\right\}\nonum \\
=& \left[(A-1) \left( 1 + (B+1)g^F_Q+(X+1)g^S_Q\right)\right]/g^E_Q + \log A  +  \mathcal{O}\left({g^{E}_Q}^{-2}\right).
\end{align}

When the expression in Eq.~\eqref{eq:frac2} replaces the logarithm in the denominator of Eq.~\eqref{meqns_A2AEfastIslowI1}, the resulting version of the firing rate, $m_Q(t)$, can be expanded once again for large $g^E_Q$ to produce the expression 

\begin{align}\label{eq:frac3}
&  \frac{1+g^E_Q+g^F_Q+g^S_Q}{\tau \log A}  \bigg[1 - \frac{(A-1)}{\log A}\left[1  + (B+1)g^F_Q  +(X+1)g^S_Q\right]\frac{1}{g^E_Q} +  \mathcal{O}\left(g^{E-2}_Q\right)\bigg].  
\end{align}

To complete the linearization, we neglect higher order terms in Eq.~\eqref{eq:frac3} and thus arrive at a linear firing-rate equation for the superthreshold case ($V_{s,Q}>V_T$). Considering again that the firing rate is taken as vanishing in the subthreshold case, when linearized about large $g^E_Q$, Eq.~\eqref{meqns_A2AEfastIslowI1} turns into the equation

\begin{align}\label{mQ_linearized}
\tilde{m}_Q = &\left\{\frac{1}{\tau \log A}\left[  \left(1- \frac{A-1}{\log A}\right) + g^E_Q  \right.\right.\\
&\left.\left. +\left(1- \frac{(A-1)(B+1)}{\log A}\right)g^F_Q  +\left(1- \frac{(A-1)(X+1)}{\log A}\right)g^S_Q\right]\right\}^+,
\end{align}
where, again, $\{x \}^+ = x $ if $x>0$, and zero otherwise. The inequality condition for $\tilde{m}_Q$ not to vanish, similarly to the condition for the nonlinear firing rate in Eq.~\eqref{boundarycondition2again_a}, is given by the inequality

\begin{align}\label{boundarycondition3}
& g^E_Q +\left(1- \frac{(A-1)(B+1)}{\log A}\right)g^F_Q  +\left(1- \frac{(A-1)(X+1)}{\log A}\right)g^S_Q+ \left(1- \frac{A-1}{\log A}\right) > 0.
\end{align}

Using the relations in Eqs.~\eqref{eqn:substitutions_a}, the piecewise-linear, per-neuron firing rate for each excitatory neuron (taking $Q = E$) becomes a function of the effective conductance variables, $g^E$, $g^F$, and $g^S$:

\begin{align}\label{mEfunctionofgEgFgS}
\tilde{m}_E =&\left\{ \frac{1}{\tau \log A}\left[ \left(\spee g^E+f_E\nu\right)   + 1- \frac{A-1}{\log A} +\left(1- \frac{(A-1)(B+1)}{\log A}\right)\speF g^F\right.\right. \nonum \\
&\left. \left.+\left(1- \frac{(A-1)(X+1)}{\log A}\right) \speS g^S \right]\right\}^+. 
\end{align}

Likewise, taking $Q = I$, the piecewise-linear, per-neuron, inhibitory firing rate is defined as  

\begin{align}
\tilde{m}_I =& \left\{\frac{1}{\tau \log A}\left[  \left(1-\frac{(A-1)(B+1)}{\log A}\right) \spiF g^F+ \left(1-\frac{(A-1)(X+1)}{\log A}\right) \spiS g^S \right. \right.\nonum\\
&+ \left.\left. \spei g^E+f_I\nu +1 - \frac{A-1}{\log A} \right]\right\}^+.\label{mIfunctionofgEgFgS}
\end{align}

The linearized, four-dimensional system for the derivatives of the effective conductance variables, incorporating Eqs.~\eqref{mEfunctionofgEgFgS} and \eqref{mIfunctionofgEgFgS}, is 

 \begin{subequations}\label{linearized_4Dsystem1}
\begin{align}
 \sigE\frac{dg^E}{dt} &= -g^E + \tilde{m}_E(t),    \label{gE_pw_linear_A2AfastslowI1} \\
\sigF\frac{dg^F}{dt} & = -g^F +  \tilde{m}_I(t),     \label{gF_pw_linear_A2AfastslowI1} \\
\sigS\frac{dg^S}{dt} &= -g^S + h,   \label{gS_pw_linear_A2AfastslowI1}\\
\rho_{\red{S}}\frac{dh}{dt} &= -h +  \tilde{m}_I(t). \label{h_pw_linear_A2AfastslowI1}
\end{align}
\end{subequations}
As the only nonlinearities in Eqs.~\eqref{4Dsystem} appear in the firing-rate terms, Eqs.~\eqref{linearized_4Dsystem1} are piecewise-linear due to the piecewise-linear nature of the linearized firing rates $\tilde{m}_E$ and $\tilde{m}_I$. The conditions upon which different parts of the piecewise-linear differential equations apply are thus the conditions required for $\tilde{m}_Q$ not to vanish, given in Eq.~\eqref{boundarycondition3} with the relations in Eqs.~\eqref{eqn:substitutions_a}. The three-dimensional $g^Eg^Fg^S$-space is thus divided into four regions by the two interesting planes spanned by $g^E$, $g^F$, and $g^S$, and defined by setting the conditions in Eqs.~\eqref{boundarycondition3} to zero. As solutions of Eqs.~\eqref{linearized_4Dsystem1} are computed, the region within which the trajectory moves dictates which parts of the piecewise-defined derivatives describe the behavior in that region.

We now comment on why we linearize the FR model by expanding about large values of the excitatory conductances.   This is because, in that regime, firing rates are also large.   At large firing rates, in turn, the response of a neuronal network (such as an I\&F model) asymptotically becomes independent of the amount of fluctuations in the dynamics; see ref.~\cite{KTRC09}, especially the results of numerical simulations shown in Fig.~8.   Additionally, the response of the linearized FR model discussed in this section turns out to resemble that of a neuronal network in the \emph{fluctuation-driven} regime.   Therefore, the linearized FR model has a convenient physiological interpretation: it provides a FR model corresponding to an I\&F model in the \emph{fluctuation-driven} regime.

\begin{figure*}[h]
\centering
      	\includegraphics[width = \textwidth]{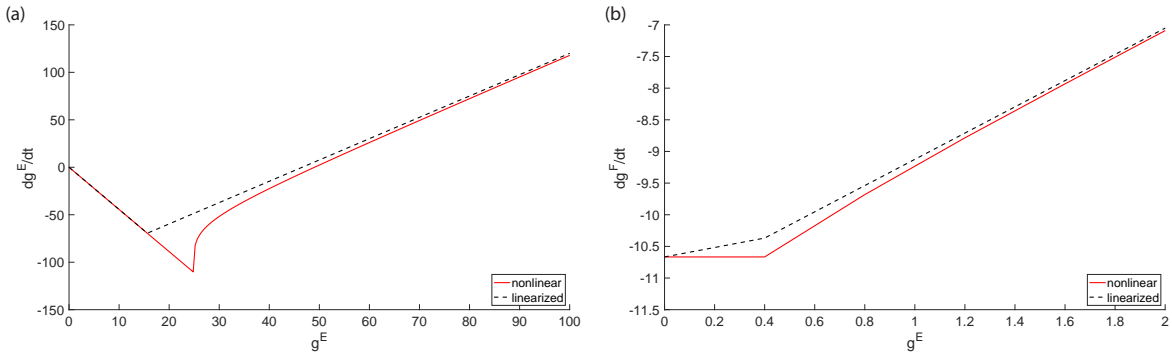} 
		\caption{The curves representing the nonlinear (solid red lines) and piecewise-linear (dashed black lines) functions that define (a) $dg^E/dt$, in Eqs.~\eqref{4eqn_gE} and \eqref{gE_pw_linear_A2AfastslowI1} as $g^E$ varies, with $g^F = 4$, and (b) $dg^F/dt$ in Eqs.~\eqref{4eqn_gF} and \eqref{gF_pw_linear_A2AfastslowI1}, versus $g^E$ for $g^F = 9.6$. Note the agreement between the nonlinear and piecewise-linear curves for large $g^E$ in both cases. Parameters in Table~\ref{Tbl:LargeParameterList}, except for the parameters pertaining to the slow effective conductance $g^S$.   The slow effective conductance $g^S$ is fixed at  $g^S=0$.}\label{fig:EFastI_gIlinearization_withplanes}
\end{figure*}

When the slow effective conductance $g^S$ is held fixed, we can depict the nonlinear and linearized derivatives of the excitatory and fast inhibitory effective conductance variables $g^E$ and $g^F$ given in Eqs.~\eqref{4Dsystem} and \eqref{linearized_4Dsystem1}, as surfaces over the $g^E$-$g^F$-plane.   In Fig.~\ref{fig:EFastI_gIlinearization_withplanes}, we show representative slices through these surfaces, and thus compare each of the nonlinear derivatives of the effective fast conductances $g^E$ and $g^F$, shown by solid red lines, with their respective linearized derivatives, shown by dashed black lines, over an interval of $g^E$, for a fixed value of $g^F$. 
 Notice that the two sets of curves are tangential in the limit of large excitatory effective conductance $g^E$.


\subsection{Existence of a Unique Limit Cycle} \label{apdx:limitcycle}

In this section, we use an approach alternative to that presented in Section~\ref{BIFURCATIONS} in order to verify the existence of a unique limit cycle in the piecewise-linear FR model, Eqs.~\eqref{linearized_4Dsystem1}, with the effective slow conductance variable, $g^S$, held fixed at a constant value. We achieve this verification with a combination of analytical solutions within each of the above-described regions in which the piecewise-defined linearized FR model has a linear form, and numerical solves of single-variable transcendental equations at each boundary between two such regions. Lastly, we confirm the existence of a limit cycle through an iterative process of possible initial conditions and associated solution trajectories. 

With the simplification of the effective slow conductance, $g^S$, as a fixed value, the piecewise-linear system in Eq.~\eqref{linearized_4Dsystem1} becomes a two-dimensional system of the excitatory and fast inhibitory effective conductances $g^E$ and $g^F$, respectively. The variable $h$ is also no longer necessary when $g^S$ is held fixed. The two-dimensional system is given by the equations  

\begin{subequations}\label{linearized_2Dsystem}
\begin{align}
 \sigE\frac{dg^E}{dt} &= -g^E + \tilde{m}_E(t),    \label{gE_pw_linear_2D} \\
\sigF\frac{dg^F}{dt} & = -g^F +  \tilde{m}_I(t),     \label{gF_pw_linear_2D} 
\end{align}
\end{subequations}
with  $\tilde{m}_E(t)$ and $\tilde{m}_I(t)$ given in Eqs.~\eqref{mEfunctionofgEgFgS} and~\eqref{mIfunctionofgEgFgS}, respectively, for a fixed value of $g^S$. 

Additionally, the three-dimensional boundary surfaces described in Appendix~\ref{apdx:linearizedmodel} for the intact linearized FR model, become two intersecting lines in the $g^E$-$g^F$-plane given by the equations of the firing rates with the slow effective conductance $g^S$ fixed, 

\begin{subequations}\label{boundarylines_2D}
\begin{align}
\tilde{m}_E &= 0, \label{boundarylines_2D_E}\\
\tilde{m}_I &= 0, \label{boundarylines_2D_F}
\end{align}
\end{subequations}
with the firing rates $\tilde{m}_E$ and $\tilde{m}_I$ given in Eqs.~\eqref{mEfunctionofgEgFgS} and~\eqref{mIfunctionofgEgFgS}, respectively.

Within each of the four regions determined by the above boundary lines, the system in Eqs.~\eqref{linearized_2Dsystem} is described by a different subset of parts of the piecewise-defined functions on its right-hand sides. We solve each system explicitly up to arbitrary constants of integration, determined by initial conditions. As the trajectory moves through one region and meets a boundary line, new initial conditions are determined by the intersection of the trajectory and the boundary, and the trajectory is then computed for the next region. Computing the new initial conditions as the trajectory crosses each boundary requires solving for the roots of a transcendental function, for which we employ \textsc{Matlab}'s built-in root-finding algorithm, \verb|fzero|. Proceeding in this manner through all four regions, the solution trajectory can be determined analytically, except for these four solves. 

Specifically, we consider the two-dimensional system in Eq.~\eqref{linearized_2Dsystem} with fixed slow inhibitory effective conductance $g^S$, and begin with an initial value for the fast inhibitory effective conductance, $g^F(0) = l$. For convenience, we consider that the initial condition for the solution lies on the inhibitory boundary line, $\tilde{m}_I = 0$, from Eq.~\eqref{boundarylines_2D_F}, and thus the excitatory effective conductance begins at the value given by

\begin{equation}
\begin{aligned}
g^E(0)  = &-\frac{1}{\spei} \left\{ f_I\nu+ \left(1-\tfrac{(A-1)(B+1)}{\log A}\right) \spiF l \right.\\
& \left. + \left(1-\tfrac{(A-1)(X+1)}{\log A}\right) \spiS g^S +1 - \frac{A-1}{\log A}  \right\}.
\end{aligned}\end{equation}
Starting at the above-described point, $(g^E(0), g^F(0))$, the trajectory is determined by the system of equations,

\begin{subequations}\label{firstregion_system}
\begin{align}
 \sigE\frac{dg^E}{dt} =& -g^E ,    \label{gE_linearpart} \\
\sigF\frac{dg^F}{dt}  =& -g^F +  \frac{1}{\tau \log A}\left[  \left(1-\frac{(A-1)(B+1)}{\log A}\right) \spiF g^F \right. \nonum \\
& \left. + \left(1-\frac{(A-1)(X+1)}{\log A}\right) \spiS g^S+ \spei g^E+f_I\nu +1 - \frac{A-1}{\log A} \right],   \label{gF_pw_linearizedpart} 
\end{align}
\end{subequations}
 in the first region through which it moves. The system in Eq.~\eqref{firstregion_system} has the explicit solution given by the equations
 
\begin{subequations}\label{firstregion_soution}
\begin{align}
g^E(t) =& C_1e^{-t/\sigE} ,    \label{gE_firstpart_soln} \\
g^F(t) =& C_2e^{-\kappa  t/\sigF}+  \frac{C_1\spei e^{-t/\sigE} + f_I\nu +1 - \frac{A-1}{\tau\ln A}}{\tau\left(\kappa - \frac{\sigF}{\sigE}\right)\ln A} - \frac{  f_I\nu +1 - \frac{A-1}{\tau\ln A}}{\tau \frac{\sigE}{\sigF}\left(\kappa^2 - \kappa\right)\ln A },  \label{gF_firstpart_soln} 
\end{align}
\end{subequations}
where $\kappa = \left[ 1- \frac{\spiF}{\tau\ln A} \left(1-\frac{(A-1)(B+1)}{\log A}\right)\right]$, and $C_1$ and $C_2$ are constants of integration determined by the initial condition established above. The value of $t$ at which this trajectory meets the excitatory boundary in Eq.~\eqref{boundarylines_2D_E} is determined using \textsc{Matlab}'s built-in root-finding algorithm, and the value of Eq.~\eqref{firstregion_soution} at that point becomes the initial condition for the solution calculation in the next region. Similar systems of equations and solutions to these systems can be determined explicitly in the same fashion. The trajectory moves through the second region where the linearized firing-rate equations, $\tilde{m}_E$ and $\tilde{m}_I$ are nonzero, then through the third region where $\tilde{m}_E$ is nonzero and $\tilde{m}_I = 0$, and finally moves through the fourth region where $\tilde{m}_E= \tilde{m}_I = 0$, until the trajectory reaches the first region again.  At each point where the trajectory meets a boundary line, new initial conditions are determined. Explicit systems that determine the dynamics of the solution trajectory with explicit solutions, up to constants of integration, determined later by initial conditions, for each region of the two-dimensional $g^E$-$g^F$-plane, can be written down. (See ref.~\cite{Pyzzathesis} for the complete set of systems and solutions.)   Note that the initial condition choice on the boundary line $\tilde m_I=0$ is made for convenience; for our argument below, it could be made along any of the boundaries dividing the regions in which Eqs.~\eqref{linearized_2Dsystem} are linear, or, in principle, along any sufficiently long curve segment transverse to the trajctories of the system.  

The process by which we calculate the trajectory for a given value, $g^F(0) = l$, may be repeated for a range of different values of $l$.  We then consider the value of the fast inhibitory effective conductance, $g^F(t)$, when the trajectory returns to the part of the boundary line, $\tilde{m}_I=0$, from which it originated, after one excursion away from it.   In this way, we generate a Poincar\'e map, $P(l)$, of this boundary line into itself.   Investigating $P(l)-l$ for a range of values of $l$, we find that the map $P(l)$ has a unique fixed point.   In other words, there is a particular value of $l$ such that the trajectory will begin and end at the same point, which indicates the existence of a unique limit cycle for a given parameter set. 

\begin{figure*}[h!]
\centering
         \includegraphics[width = \textwidth]{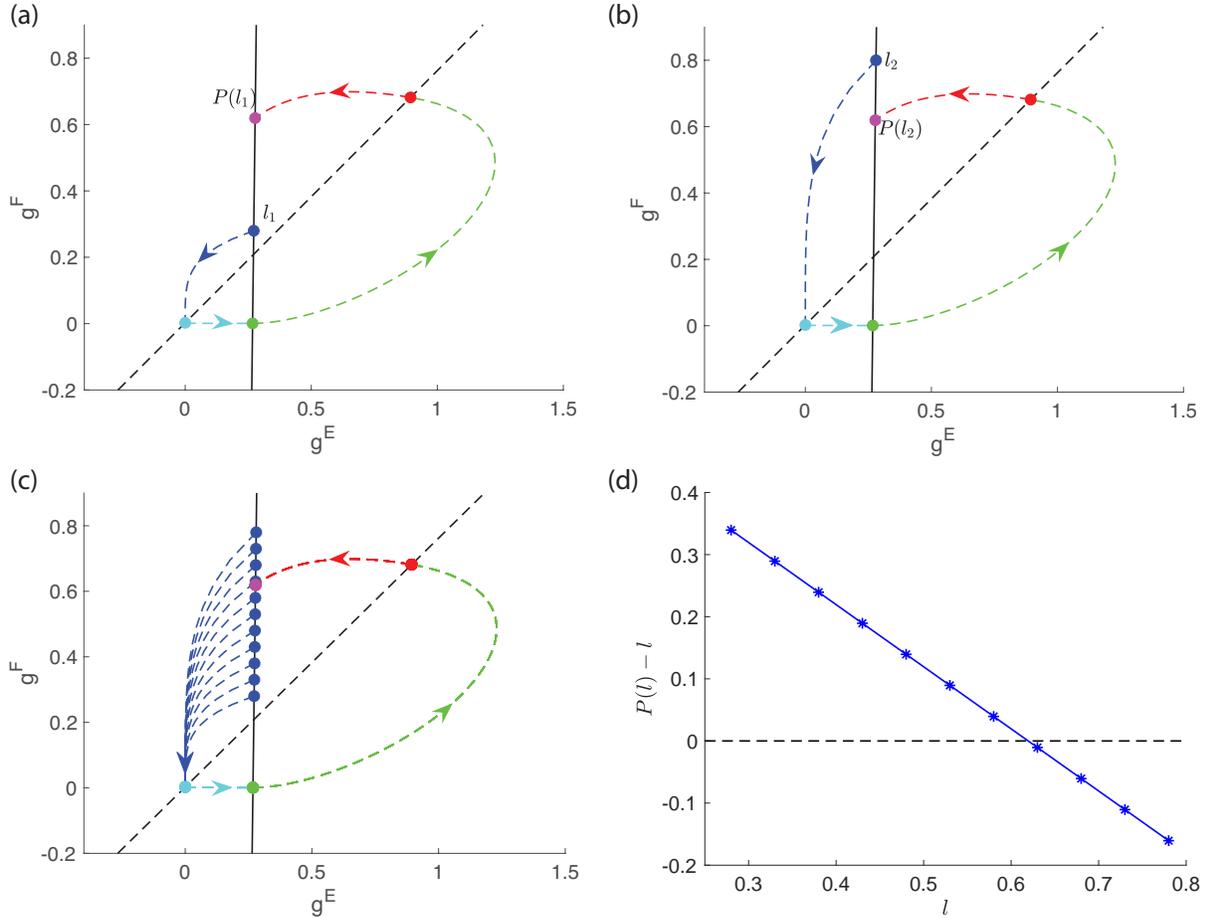}
  	 \caption{ The solid (dashed) black line represents the inhibitory (excitatory) boundary line given by $\tilde{m}_I = 0$ ($\tilde{m}_E= 0$). The colored, dashed curves illustrate the solution trajectory-segments through the four regions separated by these lines for the linearized firing rate system with $g^S$ frozen.  For each such trajectory segment, the corresponding dot indicates its initial condition, and the pink dot the point at which the trajectory returns to the initial portion of the boundary line.  Two example solution trajectories, one with $P(l_1)>l_1$ and another with $P(l_2)<l_2$,  are shown in panels (a) and (b).   The  mapping of the interval $l_1<l<l_2$ on the line $\tilde m_I=0$  along the analogous trajectory segments is shown in (c). Note that a single cyan, green, red, and pink dot is plotted for all the trajectories as they are indistinguishable at those points due to the strong contraction. In panel (d), the initial values for $g^F(0) = l$ are compared with the final values of $g^F=P(l)$ when the trajectory returns to the starting part of the boundary line.  Their difference, $P(l)-l$, is plotted.   The value $l$ of the fixed point occurs where $P(l)-l=0$, i.e., at the intersection between the curve and the dashed line. Parameters found in Table~\ref{figureparameters2}.}
 \label{fig:limitcycles} 
\end{figure*}

We illustrate this process in Fig.~\ref{fig:limitcycles}, in which the slow effective conductance is fixed at the value $g^S = 0$ (i.e., no slow inhibition), representing the case with the largest oscillations.   In Figs.~\ref{fig:limitcycles}(a) and (b), for  $l=l_1 = 0.28$ and $l=l_2 = 0.8$, respectively, we show two trajectory segments that progress counter-clockwise in the $g^E$-$g^F$-plane through the four regions determined by the boundary lines given in Eqs.~\eqref{boundarylines_2D}, until they return to the portion of the line $\tilde{m}_I=0$ above its intersection with the line $\tilde{m}_E=0$.  The blue, cyan, green, and red trajectory sub-segments are described by the solutions to the systems of equations for $g^E$ and $g^F$ given by the appropriate combinations of the piecewise portions of Eqs.~\eqref{gE_pw_linear_2D} and \eqref{gF_pw_linear_2D} in each of these four regions, respectively.  Note that for the first of two trajectories $P(l_1)>l_1$, and for the second $P(l_2)<l_2$.   In Figs.~\ref{fig:limitcycles}(c) and (d), we display the mapping of the interval $l_1<l<l_2$ along one excursion of the trajectories beginning on it back into a smaller interval inside $l_1<l<l_2$.   The corresponding trajectory segments of a number of equidistant initial points are plotted in Fig.~\ref{fig:limitcycles}(c), where we see that these trajectories are so strongly contracted towards one another that they are indistinguishable already on the line $\tilde m_E=0$.     A graph of the difference, $P(l)-l$, between the final and initial $g^F$-values along the trajectory segments emerging from the interval $l_1<l<l_2$ on the line $\tilde m_I=0$ is displayed in Fig.~\ref{fig:limitcycles}(d).   The initial condition for which this difference vanishes gives the fixed point of the mapping $l$ to $P(l)$, and thus the unique  limit cycle.


\renewcommand{\arraystretch}{1.5}
\begin{table}[h!]
\caption[Model Parameters for Numerical Limit Cycle Plots]{Parameters used for visualizing boundary lines and computing solutions to the firing-rate system in Eq.~\eqref{linearized_2Dsystem} with $g^S$ held constant in Fig.~\ref{fig:limitcycles}.}
\label{figureparameters2}
\begin{center}
\begin{tabular}{|c|c|c|c|}
\hline
\text{parameter} & \text{value} &\text{parameter} & \text{value} \\
\hline
$\nu$ &5300\text{ Hz }& $\spee $ &5\\ 
$f_E\nu$ & 6.625& $\spei$ & 4.5  \\ 
 $f_I\nu$ & 0.53& $\speF$ & 15  \\
 $\sigma_E$  & 2\text{ ms }& $\spiF $ & 0.01 \\
 $\sigma_F$  & 8\text{ ms }& $g^S$  & 0\\
 \hline
\end{tabular}
\end{center}
\end{table}

\section*{Acknowledgements}
The authors would like to thank Jennifer Crodelle, Mainak Patel, and Steven Epstein  for useful discussions.    GK, KN, PP, and DZ dedicate this paper to our late coauthor and mentor DC.

\section*{Declarations}

\subsection*{Funding}
This work was supported by National Key R\&D Program of China (2019YFA0709503), National Science Foundation in China with Grants No. 11671259, No. 11722107, SJTU-UM Collaborative Research Program, and the Student Innovation Center at Shanghai Jiao Tong University (DZ). This work was supported in part by the Department of Defense (DoD) through the National Defense Science \& Engineering Graduate Fellowship (NDSEG) Program, the National Science Foundation (NSF) Graduate Research Fellowship under Grant No. DGE-1247271,  and Research Training Groups under Grant No. DMS-1344962, the U.S. Department of Education through a Graduate Assistance in Areas of National Need (GAANN) grant, and Ohio Wesleyan's Thomas E. Wenzlau Grant (PP).

\subsection*{Conflict of Interest Statement}

The authors declare that the research was conducted in the absence of any commercial or financial relationships that could be construed as a potential conflict of interest.

\subsection*{Data and Code Availability}
All datasets analyzed for this study are included in the manuscript. Computer codes and raw data are available upon request.

%
%


\begin{thebibliography}{10}
\providecommand{\url}[1]{{#1}}
\providecommand{\urlprefix}{URL }
\expandafter\ifx\csname urlstyle\endcsname\relax
  \providecommand{\doi}[1]{DOI~\discretionary{}{}{}#1}\else
  \providecommand{\doi}{DOI~\discretionary{}{}{}\begingroup
  \urlstyle{rm}\Url}\fi

\bibitem{Ache:2005uq}
Ache, B.W., Young, J.M.: Olfaction: diverse species, conserved principles.
\newblock Neuron \textbf{48}(3), 417--430 (2005).
\newblock \doi{10.1016/j.neuron.2005.10.022}

\bibitem{Barbara:2005rw}
Barbara, G.S., Zube, C., Rybak, J., Gauthier, M., Grunewald, B.: Acetylcholine,
  gaba and glutamate induce ionic currents in cultured antennal lobe neurons of
  the honeybee, apis mellifera.
\newblock J. Comp. Physiol. A Neuroethol. Sens. Neural. Behav. Physiol.
  \textbf{191}(9), 823--836 (2005).
\newblock \doi{10.1007/s00359-005-0007-3}

\bibitem{Bazhenov:2001ai}
Bazhenov, M., Stopfer, M., Rabinovich, M., Abarbanel, H.D., Sejnowski, T.J.,
  Laurent, G.: Model of cellular and network mechanisms for odor-evoked
  temporal patterning in the locust antennal lobe.
\newblock Neuron \textbf{30}(2), 569--581 (2001)

\bibitem{Bazhenov:2001pd}
Bazhenov, M., Stopfer, M., Rabinovich, M., Huerta, R., Abarbanel, H.D.,
  Sejnowski, T.J., Laurent, G.: Model of transient oscillatory synchronization
  in the locust antennal lobe.
\newblock Neuron \textbf{30}(2), 553--567 (2001)

\bibitem{Borgers:2005tv}
Borgers, C., Kopell, N.: Effects of noisy drive on rhythms in networks of
  excitatory and inhibitory neurons.
\newblock Neural Comput. \textbf{17}(3), 557--608 (2005).
\newblock \doi{10.1162/0899766053019908}

\bibitem{Burkitt06a}
Burkitt, A.N.: A review of the integrate-and-fire neuron model: I. homogeneous
  synaptic input.
\newblock Biol. Cybern. \textbf{95}(1), 1--19 (2006).
\newblock \doi{10.1007/s00422-006-0068-6}

\bibitem{Burkitt06b}
Burkitt, A.N.: A review of the integrate-and-fire neuron model: {II}.
  inhomogeneous synaptic input and network properties.
\newblock Biol. Cybern. \textbf{95}(2), 97--112 (2006).
\newblock \doi{10.1007/s00422-006-0082-8}

\bibitem{cai2006}
Cai, D., Tao, L., Rangan, A.V., McLaughlin, D.W.: Kinetic theory for neuronal
  network dynamics.
\newblock Commun. Math. Sci. \textbf{4}(1), 97--127 (2006).
\newblock \urlprefix\url{https://projecteuclid.org:443/euclid.cms/1145905939}

\bibitem{Carcaud2016}
Carcaud, J., Giurfa, M., Sandoz, J.C.: Parallel olfactory processing in the
  honey bee brain: Odor learning and generalization under selective lesion of a
  projection neuron tract.
\newblock Front. Integrative Neuroscience \textbf{9}(75) (2016)

\bibitem{Cayre1999}
Cayre, M., Buckingham, S.D., Yagodin, S., Sattelle, D.B.: Cultured insect
  mushroom body neurons express functional receptors for acetylcholine, gaba,
  glutamate, octopamine, and dopamine.
\newblock Journal of Neurophysiology \textbf{81}(1), 1--14 (1999).
\newblock \doi{10.1152/jn.1999.81.1.1}.
\newblock \urlprefix\url{https://doi.org/10.1152/jn.1999.81.1.1}.
\newblock PMID: 9914262

\bibitem{Chen2005}
Chen, W.R., Shepherd, G.M.: The olfactory glomerulus: A cortical module with
  specific functions.
\newblock Journal of Neurocytology \textbf{34}(3), 353--360 (2005).
\newblock \doi{10.1007/s11068-005-8362-0}.
\newblock \urlprefix\url{https://doi.org/10.1007/s11068-005-8362-0}

\bibitem{Christensen:2000fu}
Christensen, T.A., Pawlowski, V.M., Lei, H., Hildebrand, J.G.: Multi-unit
  recordings reveal context-dependent modulation of synchrony in odor-specific
  neural ensembles.
\newblock Nat. Neurosci. \textbf{3}(9), 927--931 (2000).
\newblock \doi{10.1038/78840}

\bibitem{LECORRONC2002419}
Corronc, H.L., Alix, P., Hue, B.: Differential sensitivity of two insect
  gaba-gated chloride channels to dieldrin, fipronil and picrotoxinin.
\newblock Journal of Insect Physiology \textbf{48}(4), 419 -- 431 (2002).
\newblock \doi{https://doi.org/10.1016/S0022-1910(02)00061-6}.
\newblock
  \urlprefix\url{http://www.sciencedirect.com/science/article/pii/S0022191002000616}

\bibitem{Eisthen:2002fk}
Eisthen, H.L.: Why are olfactory systems of different animals so similar?
\newblock Brain Behav. Evol. \textbf{59}(5--6), 273--293 (2002)

\bibitem{Enell2007}
Enell, L., Hamasaka, Y., Kolodziejczyk, A., N{\"a}ssel, D.R.:
  $\gamma$-aminobutyric acid ({GABA}) signaling components in {D}rosophila:
  {I}mmunocytochemical localization of {GABAB} receptors in relation to the
  {GABAA} receptor subunit {RDL} and a vesicular {GABA} transporter.
\newblock Journal of Comparative Neurology \textbf{505}(1), 18--31 (2007).
\newblock \doi{10.1002/cne.21472}.
\newblock
  \urlprefix\url{https://onlinelibrary.wiley.com/doi/abs/10.1002/cne.21472}

\bibitem{ErmentroutKopell1986}
Ermentrout, G.B., Kopell, N.: Parabolic bursting in an excitable system coupled
  with a slow oscillation.
\newblock SIAM Journal on Applied Mathematics \textbf{46}(2), 233--253 (1986).
\newblock \urlprefix\url{http://www.jstor.org/stable/2101582}

\bibitem{Friedrich:2001zr}
Friedrich, R.W., Laurent, G.: Dynamic optimization of odor representations by
  slow temporal patterning of mitral cell activity.
\newblock Science \textbf{291}(5505), 889--894 (2001).
\newblock \doi{10.1126/science.291.5505.889}

\bibitem{FFTW}
{Frigo}, M., {Johnson}, S.G.: {FFTW:} an adaptive software architecture for the
  {FFT}.
\newblock In: Proceedings of the 1998 {IEEE} International Conference on
  Acoustics, Speech and Signal Processing, ICASSP '98 (Cat. No.98CH36181),
  vol.~3, pp. 1381--1384 (1998).
\newblock \doi{10.1109/ICASSP.1998.681704}

\bibitem{gardiner2004handbook}
Gardiner, C.W.: Handbook of Stochastic Methods for Physics, Chemistry, and the
  Natural Sciences, 3rd edn.
\newblock Springer Series in Synergetics. Springer, Germany (2004)

\bibitem{Gascuel:1991oz}
Gascuel, J., Masson, C.: A quantitative ultrastructural study of the honeybee
  antennal lobe.
\newblock Tissue Cell \textbf{23}(3), 341--355 (1991)

\bibitem{Grunewald:2003fu}
Grunewald, B.: Differential expression of voltage-sensitive {K$^+$} and
  {Ca$^{2+}$} currents in neurons of the honeybee olfactory pathway.
\newblock J Exp Biol \textbf{206}(Pt 1), 117--129 (2003)

\bibitem{Heinbockel:1998ys}
Heinbockel, T., Kloppenburg, P., Hildebrand, J.G.: Pheromone-evoked potentials
  and oscillations in the antennal lobes of the sphinx moth manduca sexta.
\newblock J. Comp. Physiol. A \textbf{182}(6), 703--714 (1998)

\bibitem{Hildebrand:1997kx}
Hildebrand, J.G., Shepherd, G.M.: Mechanisms of olfactory discrimination:
  converging evidence for common principles across phyla.
\newblock Ann. Rev. Neurosci. \textbf{20}, 595--631 (1997).
\newblock \doi{10.1146/annurev.neuro.20.1.595}

\bibitem{Izhikevich:2007p75}
Izhikevich, E.: Dynamical systems in neuroscience.
\newblock MIT Press p. 111 (2007)

\bibitem{Joerges:1997hc}
Joerges, J., Kuttner, A., Galizia, C.G., Menzel, R.: Representations of odours
  and odour mixtures visualized in the honeybee brain.
\newblock Nature \textbf{387}(6630), 285--288 (1997)

\bibitem{Johnston:1983uq}
Johnston, D., Brown, T.H.: Interpretation of voltage-clamp measurements in
  hippocampal neurons.
\newblock J Neurophysiol \textbf{50}(2), 464--486 (1983)

\bibitem{PCA}
Jolliffe, I.: Principal Component Analysis.
\newblock Springer-Verlag New York (2002)

\bibitem{Kashiwadani:1999}
Kashiwadani, H., Sasaki, Y.F., Uchida, N., Mori, K.: Synchronized oscillatory
  discharges of mitral/tufted cells with different molecular receptive ranges
  in the rabbit olfactory bulb.
\newblock Journal of Neurophysiology \textbf{82}(4), 1786--1792 (1999).
\newblock \doi{10.1152/jn.1999.82.4.1786}.
\newblock \urlprefix\url{https://doi.org/10.1152/jn.1999.82.4.1786}.
\newblock PMID: 10515968

\bibitem{KAY2015}
Kay, L.M.: Olfactory system oscillations across phyla.
\newblock Current Opinion in Neurobiology \textbf{31}, 141 -- 147 (2015).
\newblock \doi{https://doi.org/10.1016/j.conb.2014.10.004}.
\newblock
  \urlprefix\url{http://www.sciencedirect.com/science/article/pii/S0959438814002049}.
\newblock SI: Brain rhythms and dynamic coordination

\bibitem{Kay:2009bh}
Kay, L.M., Beshel, J., Brea, J., Martin, C., Rojas-L{\'\i}bano, D., Kopell, N.:
  Olfactory oscillations: the what, how and what for.
\newblock Trends Neurosci. \textbf{32}(4), 207--214 (2009).
\newblock \doi{10.1016/j.tins.2008.11.008}

\bibitem{Kay:2006ea}
Kay, L.M., Stopfer, M.: Information processing in the olfactory systems of
  insects and vertebrates.
\newblock Sem. Cell Dev. Biol. \textbf{17}(4), 433--442 (2006).
\newblock \doi{10.1016/j.semcdb.2006.04.012}

\bibitem{Koch99}
Koch, C.: Biophysics of Computation.
\newblock Oxford University Press, Oxford (1999)

\bibitem{KTRC09}
Kova\v{c}i\v{c}, G., Tao, L., Rangan, A.V., Cai, D.: Fokker-{P}lanck
  description of conductance-based integrate-and-fire neuronal networks.
\newblock Physical Review E \textbf{80}, 021904 (2009)

\bibitem{Laissue:2008pd}
Laissue, P.P., Vosshall, L.B.: Brain Development in Drosophila Melanogaster,
  \emph{Advances in Experimental Medicine and Biology}, vol. 628, chap.~7, pp.
  102--114.
\newblock Springer, New York (2010).
\newblock \doi{10.1007/978-0-387-78261-4{\_}7}

\bibitem{Laurent:1996bh}
Laurent, G.: Dynamical representation of odors by oscillating and evolving
  neural assemblies.
\newblock Trends Neurosci. \textbf{19}(11), 489--496 (1996).
\newblock \doi{10.1016/S0166-2236(96)10054-0}

\bibitem{Laurent:1994fu}
Laurent, G., Davidowitz, H.: Encoding of olfactory information with oscillating
  neural assemblies.
\newblock Science \textbf{265}(5180), 1872--1875 (1994).
\newblock \doi{10.1126/science.265.5180.1872}

\bibitem{Laurent:1994dz}
Laurent, G., Naraghi, M.: Odorant-induced oscillations in the mushroom bodies
  of the locust.
\newblock J Neurosci \textbf{14}(5 Pt 2), 2993--3004 (1994)

\bibitem{Laurent:1993ye}
Laurent, G., Seymour-Laurent, K.J., Johnson, K.: Dendritic excitability and a
  voltage-gated calcium current in locust nonspiking local interneurons.
\newblock J. Neurophysiol. \textbf{69}(5), 1484--1498 (1993)

\bibitem{Laurent:2001tg}
Laurent, G., Stopfer, M., Friedrich, R.W., Rabinovich, M.I., Volkovskii, A.,
  Abarbanel, H.D.: Odor encoding as an active, dynamical process: experiments,
  computation, and theory.
\newblock Ann. Rev. Neurosci. \textbf{24}, 263--297 (2001).
\newblock \doi{10.1146/annurev.neuro.24.1.263}

\bibitem{Laurent:1996fv}
Laurent, G., Wehr, M., Davidowitz, H.: Temporal representations of odors in an
  olfactory network.
\newblock J. Neurosci. \textbf{16}(12), 3837--3847 (1996)

\bibitem{Laurent:1996lq}
Laurent, G., Wehr, M., Davidowitz, H.: Temporal representations of odors in an
  olfactory network.
\newblock J Neurosci \textbf{16}(12), 3837--3847 (1996)

\bibitem{Lei:2002bv}
Lei, H., Christensen, T.A., Hildebrand, J.G.: Local inhibition modulates
  odor-evoked synchronization of glomerulus-specific output neurons.
\newblock Nat. Neurosci. \textbf{5}(6), 557--565 (2002).
\newblock \doi{10.1038/nn859}

\bibitem{Lei11108}
Lei, H., Christensen, T.A., Hildebrand, J.G.: Spatial and temporal organization
  of ensemble representations for different odor classes in the moth antennal
  lobe.
\newblock Journal of Neuroscience \textbf{24}(49), 11108--11119 (2004).
\newblock \doi{10.1523/JNEUROSCI.3677-04.2004}.
\newblock \urlprefix\url{http://www.jneurosci.org/content/24/49/11108}

\bibitem{Rangan2016}
Lei, H., Yu, Y., Zhu, S., Rangan, A.V.: Intrinsic and network mechanisms
  constrain neural synchrony in the moth antennal lobe.
\newblock Frontiers in Physiology \textbf{7}, 80 (2016).
\newblock \doi{10.3389/fphys.2016.00080}.
\newblock
  \urlprefix\url{https://www.frontiersin.org/article/10.3389/fphys.2016.00080}

\bibitem{Leitch:1996ly}
Leitch, B., Laurent, G.: Gabaergic synapses in the antennal lobe and mushroom
  body of the locust olfactory system.
\newblock J. Comp. Neurol. \textbf{372}(4), 487--514 (1996).
\newblock
  \doi{10.1002/(SICI)1096-9861(19960902)372:4{$<$}487::AID-CNE1{$>$}3.0.CO;2-0}

\bibitem{MacLeod:1998rr}
MacLeod, K., Backer, A., Laurent, G.: Who reads temporal information contained
  across synchronized and oscillatory spike trains?
\newblock Nature \textbf{395}(6703), 693--698 (1998).
\newblock \doi{10.1038/27201}

\bibitem{MacLeod:1996th}
MacLeod, K., Laurent, G.: Distinct mechanisms for synchronization and temporal
  patterning of odor-encoding neural assemblies.
\newblock Science \textbf{274}(5289), 976--979 (1996)

\bibitem{Malsburg:1999fe}
von~der Malsburg, C.: The what and why of binding: the modeler's perspective.
\newblock Neuron \textbf{24}(1), 95--104 (1999)

\bibitem{Martinez:2008pi}
Martinez, D., Montejo, N.: A model of stimulus-specific neural assemblies in
  the insect antennal lobe.
\newblock PLoS Comput. Biol. \textbf{4}(8), e1000139 (2008).
\newblock \doi{10.1371/journal.pcbi.1000139}

\bibitem{Mazor:2005eu}
Mazor, O., Laurent, G.: Transient dynamics versus fixed points in odor
  representations by locust antennal lobe projection neurons.
\newblock Neuron \textbf{48}(4), 661--673 (2005).
\newblock \doi{10.1016/j.neuron.2005.09.032}

\bibitem{MSSW00}
McLaughlin, D., Shapley, R., Shelley, M., Wielaard, J.: A neuronal network
  model of macaque primary visual cortex ({V1}): {O}rientation selectivity and
  dynamics in the input layer $4{C}\alpha$.
\newblock Proc. Natl. Acad. Sci. USA \textbf{97}, 8087--8092 (2000)

\bibitem{Ng:2002ij}
Ng, M., Roorda, R.D., Lima, S.Q., Zemelman, B.V., Morcillo, P., Miesenbock, G.:
  Transmission of olfactory information between three populations of neurons in
  the antennal lobe of the fly.
\newblock Neuron \textbf{36}(3), 463--474 (2002)

\bibitem{PRC09}
Patel, M., Rangan, A.V., Cai, D.: A large-scale model of the locust antennal
  lobe.
\newblock J. Comp. Neurosci. \textbf{27}(3), 553--567 (2009).
\newblock \doi{10.1007/s10827-009-0169-z}

\bibitem{PRC2011}
Patel, M.J., Rangan, A.V., Cai, D.: Coding of odors by temporal binding within
  a model network of the locust antennal lobe.
\newblock Front. Comput. Neurosci. \textbf{7}, 50 (2013).
\newblock \doi{10.3389/fncom.2013.00050}

\bibitem{Perez-Orive:2002nx}
Perez-Orive, J., Mazor, O., Turner, G.C., Cassenaer, S., Wilson, R.I., Laurent,
  G.: Oscillations and sparsening of odor representations in the mushroom body.
\newblock Science \textbf{297}(5580), 359--365 (2002).
\newblock \doi{10.1126/science.1070502}

\bibitem{Pyzzathesis}
Pyzza, P.B.: Idealized models of insect olfaction.
\newblock Rensselaer Polytechnic Institute, Troy, NY (2015)

\bibitem{Rangan:2012kx}
Rangan, A.V.: Functional roles for synaptic-depression within a model of the
  fly antennal lobe.
\newblock PLoS Comput. Biol. \textbf{8}(8), e1002622 (2012).
\newblock \doi{10.1371/journal.pcbi.1002622}

\bibitem{RTKC09b}
Rangan, A.V., Tao, L., Kova\v{c}i\v{c}, G., Cai, D.: Large-scale computational
  modeling of the primary visual cortex.
\newblock In: K.~Josi\'{c}, M.~Matias, R.~Romo, J.~Rubin (eds.) Coherent
  Behavior in Neuronal Networks, \emph{Springer Series in Computational
  Neuroscience}, vol.~3, pp. 263--296. Springer-Verlag (2009)

\bibitem{RojasLbano2008OlfactorySG}
Rojas-L{\'\i}bano, D., Kay, L.M.: Olfactory system gamma oscillations: the
  physiological dissection of a cognitive neural system.
\newblock Cognitive Neurodynamics \textbf{2}, 179--194 (2008)

\bibitem{Sachse:2002aa}
Sachse, S., Galizia, C.G.: Role of inhibition for temporal and spatial odor
  representation in olfactory output neurons: a calcium imaging study.
\newblock J Neurophysiol \textbf{87}(2), 1106--1117 (2002).
\newblock \doi{10.1152/jn.00325.2001}

\bibitem{Sato:2009zr}
Sato, K., Touhara, K.: Insect olfaction: receptors, signal transduction, and
  behavior.
\newblock Results Probl Cell Differ \textbf{47}, 121--138 (2009).
\newblock \doi{10.1007/400{\_}2008{\_}10}

\bibitem{SM02}
Shelley, M., McLaughlin, D.: Coarse-grained reduction and analysis of a network
  model of cortical response. {I}. drifting grating stimuli.
\newblock J. Comp. Neurosci. \textbf{12}(2), 97--122 (2002)

\bibitem{SMSW02}
Shelley, M., McLaughlin, D., Shapley, R., Wielaard, J.: States of high
  conductance in a large-scale model of the visual cortex.
\newblock J. Comp. Neurosci. \textbf{13}(2), 93--109 (2002)

\bibitem{ST01}
Shelley, M., Tao, L.: Efficient and accurate time-stepping schemes for
  integrate-and-fire neuronal networks.
\newblock J. Comput. Neurosci. \textbf{11}(2), 111--119 (2001)

\bibitem{SG95}
Singer, W., Gray, C.: Visual feature integration and the temporal correlation
  hypothesis.
\newblock Annu. Rev. Neurosci. \textbf{18}, 555--586 (1995)

\bibitem{Sivan:2004bs}
Sivan, E., Kopell, N.: Mechanism and circuitry for clustering and fine
  discrimination of odors in insects.
\newblock Proc. Nat. Acad. Sci. U.S.A. \textbf{101}(51), 17861--17866 (2004).
\newblock \doi{10.1073/pnas.0407858101}

\bibitem{Sivan:2006fy}
Sivan, E., Kopell, N.: Oscillations and slow patterning in the antennal lobe.
\newblock J. Comput. Neurosci. \textbf{20}(1), 85--96 (2006).
\newblock \doi{10.1007/s10827-006-4087-z}

\bibitem{SNS95}
Somers, D., Nelson, S., Sur, M.: An emergent model of orientation selectivity
  in cat visual cortical simple cells.
\newblock Journal of Neuroscience \textbf{15}, 5448--5465 (1995)

\bibitem{Stopfer:1997ve}
Stopfer, M., Bhagavan, S., Smith, B.H., Laurent, G.: Impaired odour
  discrimination on desynchronization of odour-encoding neural assemblies.
\newblock Nature \textbf{390}(6655), 70--74 (1997).
\newblock \doi{10.1038/36335}

\bibitem{Stopfer:2003kl}
Stopfer, M., Jayaraman, V., Laurent, G.: Intensity versus identity coding in an
  olfactory system.
\newblock Neuron \textbf{39}(6), 991--1004 (2003)

\bibitem{STRAUSFELD1999634}
Strausfeld, N.J., Hildebrand, J.G.: Olfactory systems: common design, uncommon
  origins?
\newblock Current Opinion in Neurobiology \textbf{9}(5), 634 -- 639 (1999).
\newblock \doi{https://doi.org/10.1016/S0959-4388(99)00019-7}.
\newblock
  \urlprefix\url{http://www.sciencedirect.com/science/article/pii/S0959438899000197}

\bibitem{strogatz:2000}
Strogatz, S.H.: Nonlinear Dynamics and Chaos: With Applications to Physics,
  Biology, Chemistry and Engineering.
\newblock Westview Press (2000)

\bibitem{Tanaka:2009uq}
Tanaka, N.K., Ito, K., Stopfer, M.: Odor-evoked neural oscillations in
  drosophila are mediated by widely branching interneurons.
\newblock J. Neurosci. \textbf{29}(26), 8595--8603 (2009).
\newblock \doi{10.1523/JNEUROSCI.1455-09.2009}

\bibitem{Tre93}
Treves, A.: Mean field analysis of neuronal spike dynamics.
\newblock Network \textbf{4}(3), 259--284 (1993)

\bibitem{TKPM98}
Troyer, T., Krukowski, A., Priebe, N., Miller, K.: Contrast invariant
  orientation tuning in cat visual cortex with feedforward tuning and
  correlation based intracortical connectivity.
\newblock J. Neurosci. \textbf{18}, 5908--5927 (1998)

\bibitem{Turner2008}
Turner, G.C., Bazhenov, M., Laurent, G.: Olfactory representations by
  drosophila mushroom body neurons.
\newblock Journal of Neurophysiology \textbf{99}(2), 734--746 (2008).
\newblock \doi{10.1152/jn.01283.2007}.
\newblock \urlprefix\url{https://doi.org/10.1152/jn.01283.2007}.
\newblock PMID: 18094099

\bibitem{vetterling2002numerical}
Vetterling, W.: Numerical Recipes Example Book (C++).
\newblock Cambridge University Press (2002)

\bibitem{WACHOWIAK2006411}
Wachowiak, M., Shipley, M.T.: Coding and synaptic processing of sensory
  information in the glomerular layer of the olfactory bulb.
\newblock Seminars in Cell \& Developmental Biology \textbf{17}(4), 411 -- 423
  (2006).
\newblock \doi{https://doi.org/10.1016/j.semcdb.2006.04.007}.
\newblock
  \urlprefix\url{http://www.sciencedirect.com/science/article/pii/S1084952106000486}.
\newblock Olfaction Animal Stem Cell Types

\bibitem{Hidehiro2017}
Watanabe, H., Nishino, H., Mizunami, M., Yokohari, F.: Two parallel olfactory
  pathways for processing general odors in a cockroach.
\newblock Frontiers in Neural Circuits \textbf{11}, 32 (2017).
\newblock \doi{10.3389/fncir.2017.00032}.
\newblock
  \urlprefix\url{https://www.frontiersin.org/article/10.3389/fncir.2017.00032}

\bibitem{Watts:1998aa}
Watts, D.J., Strogatz, S.H.: Collective dynamics of `small-world'networks.
\newblock Nature \textbf{393}(6684), 440--442 (1998).
\newblock \doi{10.1038/30918}.
\newblock \urlprefix\url{https://doi.org/10.1038/30918}

\bibitem{Wehr:1996xr}
Wehr, M., Laurent, G.: Odour encoding by temporal sequences of firing in
  oscillating neural assemblies.
\newblock Nature \textbf{384}(6605), 162--166 (1996).
\newblock \doi{10.1038/384162a0}

\bibitem{WHITTINGTON2000}
Whittington, M., Traub, R., Kopell, N., Ermentrout, B., Buhl, E.:
  Inhibition-based rhythms: experimental and mathematical observations on
  network dynamics.
\newblock International Journal of Psychophysiology \textbf{38}(3), 315 -- 336
  (2000).
\newblock \doi{https://doi.org/10.1016/S0167-8760(00)00173-2}.
\newblock
  \urlprefix\url{http://www.sciencedirect.com/science/article/pii/S0167876000001732}

\bibitem{Wilson:2005kb}
Wilson, R.I., Laurent, G.: Role of gabaergic inhibition in shaping odor-evoked
  spatiotemporal patterns in the drosophila antennal lobe.
\newblock J Neurosci \textbf{25}(40), 9069--9079 (2005).
\newblock \doi{10.1523/JNEUROSCI.2070-05.2005}

\end{thebibliography}


\end{document}